\documentstyle{ar}
\newbox\grsign \setbox\grsign=\hbox{$>$} \newdimen\grdimen \grdimen=\ht\grsign
\newbox\simlessbox \newbox\simgreatbox
\setbox\simgreatbox=\hbox{\raise.5ex\hbox{$>$}\llap
     {\lower.5ex\hbox{$\sim$}}}\ht1=\grdimen\dp1=0pt
\setbox\simlessbox=\hbox{\raise.5ex\hbox{$<$}\llap
     {\lower.5ex\hbox{$\sim$}}}\ht2=\grdimen\dp2=0pt

\def\simless{\mathrel{\copy\simlessbox}}
\input psfig.sty

\begin{document}
\title{X-ray Properties of Groups of Galaxies} 
\markboth{Mulchaey}{X-ray Groups}
\author{John S. Mulchaey
\affiliation{The Observatories of the Carnegie
Institution of Washington, 813 Santa Barbara St., Pasadena, CA 91101;
e-mail: mulchaey@ociw.edu}}

\begin{abstract}
ROSAT observations indicate that approximately half of all nearby
groups of galaxies contain spatially extended X-ray emission.
The radial extent of the X-ray emission is typically 
50--500 h$^{-1}_{\rm 100}$ kpc or approximately 10--50\% of the virial
radius of the group.
 Diffuse X-ray
emission is generally restricted to groups that contain at least one
early-type galaxy. X-ray spectroscopy suggests the emission mechanism
is most likely a combination of thermal bremsstrahlung and line
emission. This interpretation requires that the entire volume of
groups be filled with a hot, low-density gas known as the intragroup
medium. ROSAT and ASCA observations indicate that the temperature of
the diffuse gas in groups ranges from approximately 0.3 keV to 2 keV.
Higher temperature groups tend to follow the correlations found for
rich clusters between X-ray luminosity, temperature, and velocity
dispersion. However, groups with temperatures below approximately 1
keV appear to fall off the cluster L$_{\rm X}$--T relationship
(and possibly the L$_{\rm X}$--$\sigma$ and $\sigma$--T cluster relationships,
although evidence for these latter departures is at the present time not very
strong.)
Deviations from the cluster L$_{\rm X}$--T relationship are consistent with
preheating of the intragroup medium by an early generation of stars
and supernovae. 

There is now considerable evidence that most X-ray groups are real,
physical systems and not chance superpositions or large-scale
filaments viewed edge-on. Assuming the intragroup gas is in
hydrostatic equilibrium, X-ray observations can be used to estimate
the masses of individual systems. ROSAT observations indicate that the
typical mass of an X-ray group is $\sim$ 10$^{13}$ h$_{\rm
100}$$^{-1}$ M$_{\odot}$ out to the radius to which X-ray emission is
currently detected.
The observed baryonic masses of groups are a small fraction of the
X-ray determined masses, which implies that groups are dominated by dark
matter. On scales of the virial radius, 
the dominant baryonic component in groups is likely the intragroup
medium.
\end{abstract}

\maketitle

\section{INTRODUCTION}

Redshift surveys of the nearby universe indicate that most galaxies
occur in small groups (e.g. Holberg 1950, Humason, Mayall \& Sandage 1956,
de Vaucouleurs 1965, Materne 1979, Huchra \& Geller 1982, Geller \& Huchra
1983, Tully 1987, Nolthenius \& White 1987).
Despite diligent work in this
area over the last two decades,
the nature of poor groups is still unclear. Dynamical studies
of groups are generally hampered by small number statistics: a typical
group contains only a few luminous galaxies. For this reason, the
dynamical properties of any individual group are always rather
uncertain. In fact, many catalogued groups may not be real physical
systems at all (e.g. Hernquist et al 1995,Frederic 1995, Ramella et al 1997),
 but rather chance superpositions or large-scale
structure filaments viewed edge-on. Given the
small number of luminous galaxies in a group, the prospects
for uncovering the nature of these systems from studying the galaxies
alone seem rather bleak.

The discovery that many groups are X-ray sources has provided
considerable new insight into these important systems. X-ray
observations indicate that about half of all poor groups are luminous
X-ray sources. In many cases, the X-ray emission is extended, often
beyond the optical extent of the group. X-ray spectroscopy suggests
the emission mechanism is a combination of thermal bremsstrahlung and
line emission from highly ionized trace elements. The spatial and
spectral properties of the X-ray emission suggest the entire volume of
groups is filled with hot, low-density gas. This gas component is
referred to as the intragroup medium, in analogy to the diffuse X-ray
emitting intracluster medium found in rich clusters (e.g. Forman \& Jones
1982).

To first order, groups can be viewed as scaled-down versions of
rich clusters. Many of the fundamental properties of groups, such as
X-ray luminosity and temperature, are roughly what one expects for a
\lq\lq cluster\rq\rq \ with a velocity dispersion of several hundred
kilometers per second. However, some important physical
differences exist between groups and clusters. The velocity dispersions of
groups are comparable to the velocity dispersions of individual
galaxies. Therefore, some processes such as galaxy-galaxy merging are
much more prevalent in groups than in clusters. Other mechanisms
that are 
important in the cluster environment, such as ram-pressure stripping
and galaxy harassment, are not expected to be important in groups.  The
spectral nature of the X-ray emission is also
somewhat different in groups than
in clusters. At the typical temperature of the intracluster medium,
almost all abundant elements are fully ionized, and the X-ray emission
is dominated by a thermal bremsstrahlung continuum. At the lower
temperatures of groups, most of the trace elements retain a few atomic
electrons, and line emission dominates the observed X-ray spectrum.
Thus, while the cluster analogy is a
useful starting point, detailed studies of groups as a class are
also important. Although no strict criterion exists for separating
groups from poor clusters, for the context of this article I will
focus on systems with velocity dispersions less than about 500 km/s.

The idea that poor groups might contain diffuse hot gas dates back to
the classic Kahn \& Woltjer (1959) paper on the \lq\lq timing
mass\rq\rq \ of the Local Group. Kahn \& Woltjer (1959) found that the
mass of the Local Group far exceeded the visible stellar mass and
suggested the bulk of the missing mass was in the form of a warm, 
low-density plasma. Although it is now generally believed that the Local
Group is dominated by dark matter, Kahn \& Woltjer's estimates for the
properties of the intragroup medium are remarkably similar to more
recent estimates. More than a decade after Kahn \& Woltjer, the idea
of diffuse gas in the Local Group and other groups was revisited by
Oort (1970), Ruderman \& Spiegel (1971), Hunt \& Sciama (1972),
and Silk \& Tarter (1973).

The earliest claims for X-ray detections of groups came from the
non-imaging X-ray telescopes Uhuru, Ariel 5, and HEAO 1 in the 1970s.
Cooke et al (1978) produced a catalog (known as the 2A) of 105 bright
X-ray sources from the Leicester Sky Survey Instrument on Ariel
5. Based on positional coincidences, Cooke et al (1978) suggested the
identification of seven X-ray sources in the 2A catalog as groups of
galaxies. Subsequent observations showed that several of these X-ray
sources were variable, indicating they were actually active galaxies
within the group (Ricker et al 1978, Ward et al 1978, Griffiths et al
1979). However, several of the remaining objects in Cooke et al (1978)
were later shown to be poor clusters (Schwartz et al 1980).

X-ray studies of lower-mass systems received a major boost with the
launch of the Einstein Observatory in November 1978. Einstein
observations firmly established that some poor clusters with bright
central galaxies (i.e. MKW and AWM clusters; Morgan et al 1975, Albert
et al 1977) were X-ray sources (Kriss et al 1980, 1983, Burns et al
1981, Price et al 1991, Dell'Antonio et al 1994). The X-ray
luminosities of these poor clusters range from several times 10$^{41}$
ergs s$^{-1}$ h$_{\rm 100}$$^{-2}$ up to several times 10$^{43}$ ergs
s$^{-1}$ h$_{\rm 100}$$^{-2}$. The X-ray emission in these poor
clusters was shown to be extended (out to radii as great as 0.5
h$_{\rm 100}$$^{-1}$ Mpc) with temperatures in the range T $\sim$ 1--5
keV. Although most of these systems are somewhat richer than the typical
groups considered in this review, these Einstein observations clearly
demonstrated that diffuse X-ray emission was not restricted to rich
clusters.

Several attempts were also made to study even poorer galaxy systems 
with Einstein.
 Biermann and collaborators detected extended
 emission in two nearby elliptical-dominated groups (Biermann et al
 1982; Biermann \& Kronberg 1983). In both cases, the X-ray emission
 was centered on the dominant galaxy. For the NGC 3607 group,
 Biermann et al (1982) concluded that the X-ray emission most likely
 originated from a hot, intergalactic gas because it was extended on
 scales larger than the galaxy (Biermann et al estimate a Gaussian
 width for the X-ray emission of 4.7$'$ $\approx$ 13 h$_{\rm
 100}$$^{-1}$ kpc). From a rough fit to the X-ray spectrum, a
 temperature of $\approx$ 5 $\times$ 10$^6$ K and an X-ray luminosity
 of 2 $\times$ 10$^{40}$ h$_{\rm 100}$$^{-2}$
 ergs s$^{-1}$ was found. Following their
 discovery of X-ray emission in the NGC 3607 group, Biermann \&
 Kronberg (1983) found a similar component in the NGC 5846 group. The
 Einstein Observatory was also used to study the X-ray properties of
 compact groups. Bahcall et al (1984) studied five compact groups,
 including four from Hickson's (1982) catalog. Three of the compact
 groups were detected with Einstein. The Einstein exposure times for
 these groups were very short, resulting in only $\sim$ 20--60 net
 counts in the X-ray detected cases. Bahcall et al (1984) noted that
 the X-ray luminosities of two of the groups were of order $\sim$
 10$^{42}$ ergs s$^{-1}$ h$_{\rm 100}$$^{-2}$, much higher than the
 X-ray emission expected from the member galaxies alone. The emission
 was also extended in these two groups, and in the case of Stephan's
 Quintet, the shape of the X-ray spectrum was unlike that expected
 from individual galaxies. These X-ray properties led Bahcall et al
 (1984) to conclude that the X-ray emission likely originated in a hot
 intragroup gas in at least two of the five groups they studied.
 Thus, although it was not possible to unambiguously separate a
 diffuse component from galaxy emission with
 Einstein, there were strong indications that intragroup gas was
 likely present in some groups.

\section{X-RAY TELESCOPES}

While there were hints from Einstein observations that some groups of
galaxies might contain a hot intragroup medium, it was not until the
1990s that the presence of diffuse gas in groups was firmly
established. Group studies were aided by the launch of two important
X-ray telescopes, ROSAT (the ROentgen SATellite) and ASCA (Advanced
Satellite for Cosmology and Astrophysics). Both of these telescopes
were capable of simultaneous X-ray imaging and spectroscopy in the
energy range appropriate for poor groups. In addition, the field of
view for both telescopes was large enough that nearby groups
could effectively be studied.

\subsection{ROSAT}
ROSAT consisted of two telescopes. The X-ray telescope (Aschenbach
1988) was sensitive to photons in the energy range of 0.1--2.4 keV,
whereas the Wide Field Camera (Wells et al 1990) covered the energy
range 0.070--0.188 keV. 
The relatively high luminosity of the X-ray background combined with 
the strong effects of absorption by the Galaxy limited the 
study of diffuse 
extragalactic gas with the Wide Field Camera.
Therefore, this instrument was not useful 
for studies of groups and will not be discussed further. Two different
kinds of detectors were used with the X-ray telescope: the Position
Sensitive Proportional Counter (PSPC) and the High Resolution Imager
(HRI). ROSAT was flown with two nearly identical PSPC detectors
(Pfeffermann et al 1988). The low internal background, large field of
view, and good sensitivity to soft X-rays made the PSPC detectors
ideal for studying X-ray emission from groups. The PSPC detectors
also had modest energy resolution, allowing the spectral properties of
the X-ray emission to be studied. Although the ROSAT HRI provided higher
spatial resolution than the PSPC detectors ($\sim$ 5$''$ versus $\sim$
25$''$ for an on-axis source), the internal background of the HRI was
high enough that the low surface brightness diffuse emission
found in groups could in general not be studied with this
instrument. Therefore, most ROSAT studies of groups were performed
with the PSPC.

The ROSAT mission consisted of two main scientific phases. The first
was a six-month, all-sky survey (Voges 1993) performed with one of the
PSPC detectors (until that detector was destroyed during an accidental
pointing at the Sun in January 1991). The mean exposure time for the
all-sky survey was approximately 400 seconds.  Following the
completion of the survey, ROSAT was operated in so-called \lq\lq
pointed mode\rq\rq\ --- that is, with longer pointings at individual
targets. Typical exposure times during the pointed mode of the
mission were in the range 5000 to 25,000 seconds, or roughly 10 to 50
times longer than the all-sky survey exposures. Although the pointed
mode of the ROSAT mission lasted until early 1999, the second PSPC
detector ran out of gas in late 1994, effectively ending studies of
diffuse emission in groups.

\subsection{ASCA}

ASCA, a joint Japanese--United States effort, was launched in early
1993. ASCA consists of four identical grazing-incident X-ray
telescopes each equipped with an imaging spectrometer (Tanaka et al
1994).  The focal plane detectors are two CCD cameras (known as the
Solid-State Imaging Spectrometers, or SIS; Gendreau 1995) and two gas
scintillation imaging proportional counters (Gas Imaging Spectrometer,
or GIS; Ohashi et al 1996). The SIS detectors have superior energy
resolution, whereas the GIS detectors provide a larger field of view. The
angular resolution of ASCA is considerably worse than that of ROSAT,
with a half power diameter of approximately 3$'$. However, ASCA's
spectral resolution is much higher than that of the ROSAT PSPC
(E/$\Delta$E $\sim$ 20 for the SIS at 1.5 keV versus E/$\Delta$E $\sim$3
for the PSPC), so this instrument has primarily played a role in the study
of the spectral properties of the intragroup gas. Although 
the detectors aboard ASCA have undergone serious degradation, this mission is
expected to remain operational until sometime in the year 2000.

\section{PROPERTIES OF THE INTRAGROUP MEDIUM}

\subsection{First ROSAT Results}

The great potential of ROSAT for group studies was demonstrated in
early papers by Mulchaey et al (1993) and Ponman \& Bertram (1993).
Each of these papers presented a detailed look at the X-ray properties
of an individual group. Mulchaey et al (1993) studied the NGC 2300
group, a poor group dominated by an elliptical-spiral pair. The X-ray
emission in the NGC 2300 group is not centered on any particular
galaxy, but is instead offset from the elliptical galaxy NGC 2300 by
several arcminutes. The X-ray emission can be traced to a radius of at
least $\sim$ 150 h$_{\rm 100}$$^{-1}$ kpc ($\sim$ 25 $'$). Ponman \&
Bertram (1993) studied Hickson Compact Group 62 (HCG 62). In this
case, the X-ray emission is extended to a radius of at least 210
h$_{\rm 100}$$^{-1}$ kpc ($\sim$ 18$'$). Although the presence of
intragroup gas had been suggested by earlier Einstein observations,
these ROSAT PSPC results were the first to unambiguously separate a
diffuse component related to the group from emission associated with
individual galaxies. The intragroup medium interpretation was also
supported by the ROSAT PSPC spectra, which are well-fit by a thermal
model with a temperature of approximately 1.0 keV ($\sim$ 10$^{7}$ K).
The ROSAT PSPC spectrum of HCG 62 contained enough counts that Ponman
\& Bertram (1993) could also derive a temperature profile for the gas.
Ponman \& Bertram (1993) found evidence for cooler gas near the center
of the group, which they interpreted as evidence for a cooling flow.
Many of the X-ray properties of the NGC 2300 group and HCG 62 are
consistent with the idea of these systems being scaled-down versions of more
massive clusters.

The early ROSAT observations of groups also provided some surprises.
For both groups, the gas metallicity derived from the X-ray spectra
was much lower than the value found for rich clusters ($\sim$ 6\%
solar for NGC 2300 and $\sim$ 15\% solar for HCG 62, compared with
$\sim$ 20--30\% solar found for clusters; Fukazawa et al 1998).  
The X-ray data were also used to estimate the
total masses of the groups. In each case, the mass of the group is
approximately 10$^{13}$ h$_{\rm 100}$$^{-1}$ M$_{\odot}$. Comparing
the total mass as measured by the X-ray data with the total mass in
observed baryons, Mulchaey et al (1993) and Ponman \& Bertram (1993)
concluded that the majority of mass in these groups is
dark. In the case of the NGC 2300 group, Mulchaey et al (1993) estimated
a baryon fraction that was low enough to be consistent with
$\Omega$=1 and the baryon
fraction predicted by standard big bang nucleosynthesis. However, 
subsequent analysis of the ROSAT PSPC data
suggests the true baryon fraction is higher in this group (
Henriksen \& Mamon 1994; David et al 1995,
Pildis et al 1995, Davis et al 1996).

\subsection{ROSAT Surveys of Groups}

Unfortunately, the results of Mulchaey et al (1993) and Ponman \&
Bertram (1993) came late enough in the lifetime of the ROSAT PSPC that
large systematic follow-up surveys of groups were not carried out with
this instrument. However, the ROSAT PSPC observed many galaxies
during its lifetime, and because most galaxies occur in groups, many
groups were observed serendipitously. Furthermore, the field of view
of the PSPC was large enough that many groups were also observed when
the primary target was a star, an active galaxy, or a QSO. In the end,
over 100 nearby groups were observed by the ROSAT PSPC during its
lifetime, and most of our current understanding of the X-ray properties
of groups comes from this dataset.

The existence of an excellent data archive has led to many X-ray
surveys of groups using ROSAT PSPC data (Pildis et al 1995, David et
al 1995, Doe et al 1995, Saracco \& Ciliegi 1995, Mulchaey et al
1996a, Ponman et al 1996, Trinchieri et al 1997, Mulchaey \& Zabludoff
1998, Helsdon \& Ponman 2000; Mulchaey et al 2000). 
These surveys indicate that not all
poor groups contain an X-ray---emitting intragroup medium. The exact
fraction of groups that contain hot intragroup gas has been difficult
to quantify because of biases in the sample selection. For example,
many of the samples used in archival surveys contain groups that were a priori
known to be bright X-ray sources or were likely to be bright X-ray sources
based on morphological selection (such as a high fraction of
early-type galaxies). These samples are almost certainly not
representative of poor groups in general. Furthermore, the term
\lq\lq X-ray detected\rq\rq \ has a variable meaning in the
literature; some authors use this term only when a diffuse, extended
X-ray component (i.e. intragroup medium) is present, whereas others also
include cases when emission is associated primarily with the
individual galaxies.

There has been considerable interest in the Hickson Compact Groups
(HCGs; Hickson 1982; for a review see Hickson 1997).
The short crossing times implied for these systems has led some authors
to suggest the HCGs are chance alignments of unrelated galaxies within
looser systems (Mamon 1986, Walke \& Mamon 1989), bound configurations
within loose groups (Diaferio et al 1994, Governato et al 1996) 
or filaments viewed edge-on (Hernquist et al 1995). X-ray observations
can potentially help distinguish between these various scenarios 
(Ostriker et al 1995; Diaferio et al 1995).
Ebeling et al (1994) detected 
eleven HCGs in the ROSAT All-Sky Survey
(RASS) data. For some of the
detections, the X-ray emission was clearly extended and thus
consistent with hot intragroup gas. However, in other cases the
sensitivity of the RASS was not good enough to determine the nature of
the X-ray emission. Still, Ebeling et al's sample was the first to
suggest a correlation between the presence of X-ray emission and a
high fraction of early-type galaxies in groups. Pildis et al (1995)
and Saracco \& Ciliegi (1995) each analyzed ROSAT pointed-mode
observations of 12 HCGs (there was considerable overlap in these two
samples). Both surveys found that approximately two- thirds of the
HCGs were X-ray detected, although in many cases the X-ray emission
could not be unambiguously attributed to intragroup gas. (Note also
that many of the X-ray detections in these two surveys overlapped with
Ebeling et al's earlier RASS detections.) A much more complete study
of the HCGs was presented by Ponman et al (1996). This survey combined
pointed ROSAT PSPC observations with ROSAT All-Sky Survey data to search for
diffuse gas in 85 HCGs. These authors detected extended X-ray emission
in $\sim$ 26\% (22 of 85 groups) of the systems studied and inferred that
$\sim$ 75\% of the HCGs contain a hot intragroup medium (when one
corrects for the detection limits of the
observations). Although this is intriguing, some caution must be expressed
regarding the Ponman et al (1996) results. Given the compactness of
these groups, the nature of the X-ray emission in some of the detected
HCGs is far from clear. For example, although Stephan's Quintet (HCG
92) is extended in the ROSAT PSPC data (Sulentic et al 1995), a 
higher-resolution ROSAT HRI image suggests that most of the extended emission
is associated with a shock feature and not with a smooth intragroup
gas component (Pietsch et al 1997). Thus, some of the detections in
the Ponman et al (1996) survey may not be related to an intragroup
medium at all.

Many of the problems inherent to the study of compact groups can be
avoided with loose groups. Helsdon \& Ponman (2000) studied a sample
of 24 loose groups from the catalog of Nolthenius (1993) and found that
half of the systems contain intragroup gas.  Mulchaey et al (2000)
detected diffuse gas in 27 of 57 groups selected from redshift surveys
(including the Nolthenius catalog). Both of these studies relied on
fairly deep ROSAT pointings and therefore are sensitive to gas down to
low X-ray luminosities ($\sim$ 5 $\times$ 10$^{40}$ h$_{\rm
100}$$^{-2}$ ergs s$^{-1}$). The majority of the groups in both
Helsdon \& Ponman (2000) and Mulchaey et al (2000) were observed
serendipitously with the ROSAT PSPC. Based on their velocity
dispersions and morphological composition, these samples are fairly
representative of groups in nearby redshift surveys. Therefore, 
these surveys suggest that 
$\sim$ 50\% of nearby optically- selected groups contain a
hot X-ray--emitting intragroup medium.

ROSAT All-Sky Survey (RASS) data have also played an important role in
our understanding of the X-ray properties of groups. While the RASS
observations are generally not very deep, the nearly complete coverage
of the sky allows for larger samples to be studied than is possible with
the pointed mode data 
alone. Henry et al (1995) used the RASS data in the
region around the north ecliptic pole to define the first X-ray
selected sample of poor groups.
The survey by Henry et al (1995) was sensitive
to all groups more luminous than $\sim$ 2.3 $\times$ 10$^{41}$ h$_{\rm
100}$$^{-2}$ ergs s$^{-1}$. Although their sample was rather small (8
groups), Henry et al (1995) were able to show that X-ray--selected
groups lie on
the smooth extrapolation of the cluster X-ray luminosity and
temperature functions. The X-ray selected groups also have lower
spiral fractions than typical optically- selected groups, which
may suggest that X-ray selection produces a more dynamically evolved
sample of groups (Henry et al 1995).

The RASS data has also been used to study optically- selected group
samples. Burns and collaborators have devoted considerable effort
into studying the X-ray properties of the WBL poor clusters and
groups (White et al 1999), which were selected by galaxy surface density. 
One of the more important results from these studies is the
derivation of the first X-ray luminosity function for an optically
selected sample of groups and poor clusters (Burns et al 1996). The
luminosity function derived by Burns et al (1996) is a smooth
extrapolation of the rich cluster X-ray luminosity function and
is consistent with the luminosity function Henry et al (1995) derived
from their X-ray selected sample of groups. Follow-up work on some of
the brighter sources in the WBL catalog indicates that many of these
objects are more massive than typical groups with gas temperatures of
2--3 keV (Hwang et al 1999). These systems are important because they
represent the transition objects between poor groups and rich
clusters.

Mahdavi et al (1997, 2000) used the RASS database to study the X-ray
properties of a large sample of 
groups selected from the CfA redshift survey 
(Ramella et
al 1995). After accounting for selection effects, 
Mahdavi et al (2000) estimate that $\sim$ 40\% of the groups 
are extended X-ray sources.
From these
detections, the authors derive a relationship between X-ray luminosity
and velocity dispersion that is much shallower than is found for rich
clusters (see Section 4.2).
They suggest that this result is consistent with the
X-ray emission in low velocity dispersion groups
being dominated by intragroup gas bound to the
member galaxies as opposed to the overall group potential. 
Unfortunately, the RASS observations of
groups typically contain very few 
counts, so detailed spatial studies of the emission are not possible
with this dataset. A much deeper X-ray survey of an optically-selected 
group sample like the one used in Mahdavi et al (2000) would be very
useful and should be a priority for future X-ray missions.

\subsection{Spatial Properties of the Intragroup Medium}

\subsubsection{X-RAY MORPHOLOGIES}

The morphology of the X-ray emission can provide important clues into
the nature of the hot gas. 
There is a considerable range in the observed X-ray morphologies of
groups. X-ray luminous (L$_{\rm X}$ $>$ 10$^{42}$ h$_{\rm 100}$$^{-2}$
erg s$^{-1}$) groups tend to have somewhat
 regular morphologies (see Figure 1).
The total extent of the X-ray emission in these cases is often beyond
the optical extent of the group as defined by the galaxies. The peak
of the X-ray emission is usually coincident with a luminous elliptical
or S0 galaxy, which tends to be the most optically luminous group
member (Ebeling et al 1994, Mulchaey et al 1996a, Mulchaey \&
Zabludoff 1998). The position of the brightest galaxy is also
indistinguishable from the center of the group potential, as defined
by the mean velocity and projected spatial centroid of the group
galaxies (Zabludoff \& Mulchaey 1998). Therefore, the brightest
elliptical galaxy lies near the dynamical center of the group. There is
also a tendency for the diffuse X-ray emission to roughly align with
the optical light of the galaxy in many cases (Mulchaey et al 1996a,
Mulchaey \& Zabludoff 1998). These morphological characteristics are
similar to those found for rich clusters containing cD galaxies
(e.g.  Rhee et al 1992, Sarazin et al 1995, Allen et al 1995).

\begin{figure}
\psfig{figure=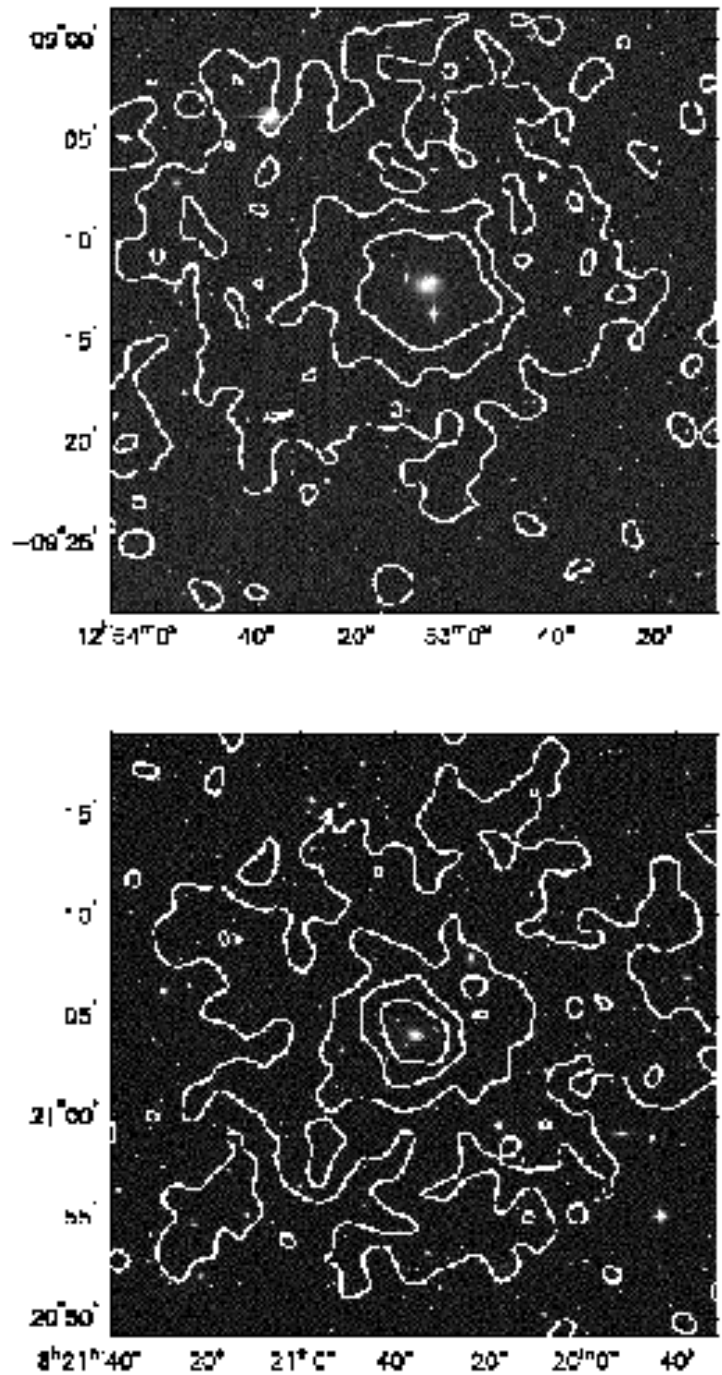,width=4in}
\caption{
Contour map of the diffuse X-ray emission as traced by the ROSAT
PSPC in HCG 62 ({\it top}) and
the NGC 2563 group ({\it bottom}) overlayed on the STScI Digitized Sky
Survey. The X-ray data have been smoothed with a Gaussian profile of
width 30$''$. The coordinate scale is for epoch J2000.}
\end{figure}

At lower luminosities, more irregular X-ray morphologies are often
found (see Figure 2). In these cases, the X-ray emission is not
centered on one particular galaxy, but rather is distributed around
several galaxies. Low X-ray luminosity groups also tend to have lower
gas temperatures. 
Dell'Antonio et al
(1994) and Mahdavi et al (1997) suggested that the 
change in X-ray morphologies at low X-ray luminosities 
indicates a change in the nature of the X-ray emission. 
They proposed a \lq\lq
mixed-emission\rq\rq\ scenario where the observed diffuse X-ray
emission originates from both a global group potential and from
intragroup gas in the potentials of individual galaxies. In this
model, the latter component becomes dominant in low-velocity
dispersion systems. This model is consistent with the fact that the 
X-ray emission is distributed near the luminous galaxies in many of the
low-luminosity systems. 
Another possible source of diffuse
X-ray emission in the low-luminosity systems might be gas
that is shock-heated
to X-ray temperatures by galaxy collisions and encounters within the
group environment. This appears to be the case in HCG 92, where the
diffuse X-ray emission comes predominantly from an intergalactic
feature also detected in radio continuum maps (Pietsch et al 1997).
Given that many of the groups with irregular X-ray morphologies are
currently experiencing strong galaxy-galaxy interactions (e.g. HCG 16,
HCG 90), shocks may be important in many cases.
Regardless of the
exact origin of the gas, the clumpy X-ray morphologies suggest that
the X-ray gas may not be virialized in these cases.

\begin{figure}
\psfig{figure=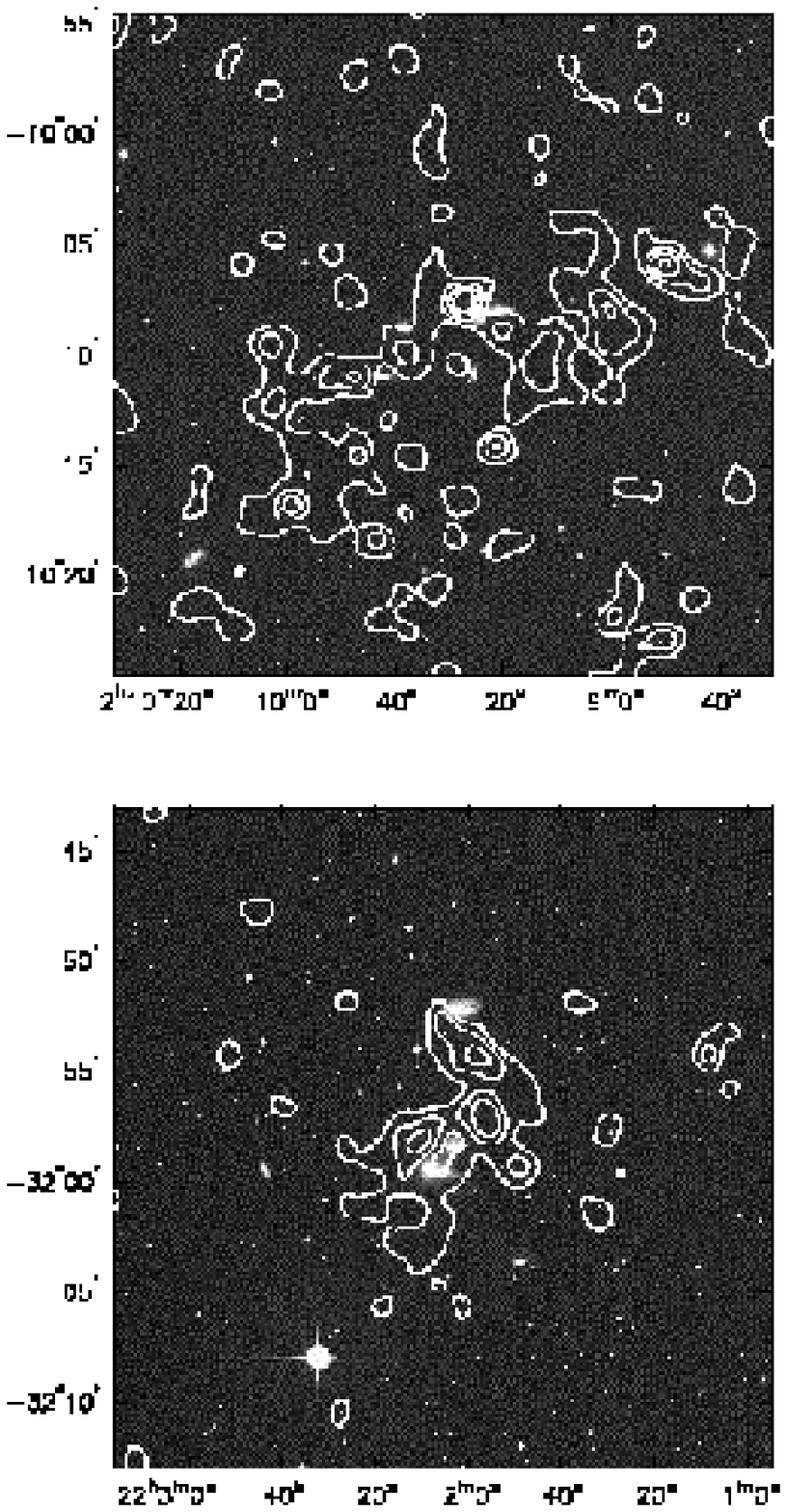,width=4.0in}
\caption{Contour map of the diffuse X-ray emission as traced by the ROSAT
PSPC in HCG 16({\it top}) and
HCG 90({\it bottom}) overlayed on the STScI Digitized Sky
Survey. The X-ray data have been smoothed with a Gaussian profile of
width 30$''$. The coordinate scale is for epoch J2000.}
\end{figure}

\subsubsection{SPATIAL EXTENT}

To estimate the extent of the hot gas, the usual method is to
construct an azimuthally-averaged surface brightness profile and 
determine at what radial distance the emission approaches the background
value. For most rich clusters, the central surface brightness of 
the intracluster medium
is several orders of magnitude higher than the surface brightness
of the X-ray background. Not surprisingly, the central surface 
brightness of less massive systems like groups tends to be much lower. 
In fact, in many of the X-ray weakest groups, the central surface
brightness of the intragroup gas is just a few times higher than 
that of the background. 
Therefore, the measured extent of the X-ray emission
in groups is usually much less than that of rich clusters. 
When comparing groups and clusters, it 
is useful to normalize the radial extent of the X-ray gas 
by the mass of the system. 
Figure 3 plots X-ray extent 
normalized by the virial radius (R$_{\rm virial}$) of each system
versus temperature for a 
sample of groups and clusters. 
Figure 3 indicates that many rich clusters are currently 
detected to approximately R$_{\rm virial}$, whereas groups are 
typically detected to a small fraction of R$_{\rm virial}$. In some
cases, the group X-ray extents are less than 10\% of the 
virial radius. There is also a strong correlation between the radius
of detection in virial units and the temperature of the gas in groups:
cool groups are detected to a smaller
fraction of their virial radius than hot groups. This correlation 
is important because it suggests that a smaller
fraction of the gas mass and thus, X-ray luminosity, is detected 
in low temperature
systems. Therefore, it is very important to account for this effect when 
one compares X-ray properties of systems spanning a large range in temperature
(i.e. mass). Unfortunately, this has generally not been done in the 
literature.

\begin{figure}
\psfig{figure=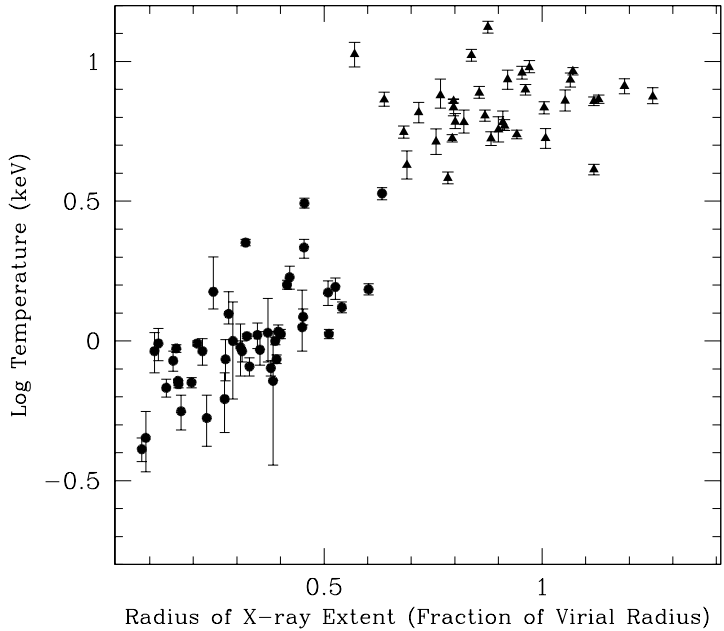,width=4.0in}
\caption{Total radius of X-ray extent plotted as a fraction of the
virial radius of each system versus the logarithm of the temperature
for a sample of 
groups (circles) and rich clusters (triangles). The groups
were taken from Mulchaey et al (1996a), Hwang et al (1999) and Helsdon \&
Ponman (2000). The clusters plotted are a redshift-selected
 subset of the clusters in White (2000). The virial radius for each system
was calculated assuming r$_{\rm virial}(T)$ =
1.85 (T/10keV)$^{0.5}$ (1+z)$^{-1.5}$
 h$_{\rm 100}^{-1}$ Mpc (Evrard et al 1996).
}
\end{figure}

\subsubsection{THE BETA MODEL}

Traditionally,
a hydrostatic isothermal model has been used to describe the
surface
brightness profiles of rich clusters (e.g. Jones \& Forman 1984). By
analogy to the richer systems, this model is usually adopted for poor
groups. The hydrostatic isothermal model assumes that both the hot gas
and the galaxies are in hydrostatic equilibrium and isothermal. These
assumptions appear to be valid for groups with regular X-ray
morphologies, but are likely incorrect for groups with irregular X-ray
morphologies (although this model is often applied even in these
cases). With King's (1962) analytic approximation to the isothermal
sphere, the X-ray surface brightness at a projected radius R is given
by:

\centerline{S(R)=S$_{\rm o}$
(1 + (R/r$_{\rm c}$)$^2$)$^{-3\beta + 0.5}$}

\noindent{where r$_{\rm c}$ is the core radius of the gas
distribution. This model is often referred to as the standard beta
model in the literature. The parameter $\beta$ is the ratio of the
specific energy in galaxies to the specific energy in the hot gas:}

\centerline{$\beta$ $\equiv$ $\mu$m$_{\rm p}$$\sigma$$^2$/kT$_{\rm
gas}$}

\noindent{where $\mu$ is the mean molecular weight, m$_{\rm p}$ is the
mass of the proton, $\sigma$ is the one-dimensional velocity
dispersion, and T$_{\rm gas}$ is the temperature of the intragroup
medium.
For high-temperature systems such as clusters, the X-ray emissivity is
fairly independent of temperature over the energy range observed by
ROSAT ($\sim$ 0.1--2 keV). Therefore, the gas density profile can be
derived from the surface brightness profile even if the gas
temperature varies somewhat within the cluster. However, at the
temperatures more typical of groups, the X-ray emissivity is a strong
function of temperature. Thus, to invert the observed surface
brightness profiles of groups to a gas density profile, the gas must
be fairly isothermal.}

Based on fits to ROSAT PSPC data, most authors have derived $\beta$
values of around $\sim$ 0.5 for groups (Ponman \& Bertram 1993, David
et al 1994, Pildis et al 1995, Henry et al 1995, Davis et al 1995,
David et al 1995, Doe et al 1995, Mulchaey et al 1996a). This number
is somewhat lower than the typical value found for clusters
(e.g. $\sim$ 0.64; Mohr et al 1999). However, simulations of clusters
indicate that the $\beta$ value derived from a surface brightness 
profile depends strongly on the 
range of radii used in the fit (Navarro et al 1995,
Bartelmann \& Steinmetz 1996). In particular, $\beta$ values derived on
scales much less than the virial radius tend to be systematically low.
As most groups are currently detected to a much smaller fraction of
the virial radius than rich clusters, a direct comparison between 
group and cluster $\beta$ values may not be particularly meaningful.

Although the hydrostatic isothermal model has almost universally been
used for groups, in most cases it provides a poor fit to the data. In
general, the central regions of groups exhibit an excess of emission
above the extrapolation of the beta model to small radii. This
steepening of the profile is often accompanied by a drop in the gas
temperature, which has led some authors to suggest that the central
deviations are related to a cooling flow (Ponman \& Bertram 1993,
David et al 1994, Helsdon \& Ponman 2000). Alternatively, the excess
flux could be emission associated with the central elliptical galaxy
(Doe et al 1995, Ikebe et al 1996, Trinchieri et al 1997, Mulchaey \&
Zabludoff 1998).

Mulchaey \& Zabludoff (1998) have shown that the surface brightness
profiles in many groups can be adequately fit using two separate beta
models. Although the various parameters are not well- constrained with
the two-component models, Mulchaey \& Zabludoff (1998) found a
systematic trend for the $\beta$ values to be larger with this model
than in the case of a single beta model. Similar behavior has been
found for rich clusters of galaxies (Ikebe et al 1996, Mohr et al
1999).  Mohr et al (1999) suggest that the effect is a consequence of the
strong coupling between the core radius (r$_{\rm c}$) and $\beta$ in
the fitting procedure; a beta model with a large core radius and
high $\beta$ value can produce a profile similar to that of a beta model
where both parameters are lower. Therefore, the presence of a central
excess drives the core radius (and thus $\beta$) to lower values in
the single beta model fits. While Helsdon \& Ponman (2000) verified
the need for multiple components in groups, they did not derive
systematically higher $\beta$ values. The likely explanation is that
the argument in Mohr et al (1999) applies exclusively to systems
where the extended component (i.e. the group/cluster gas) dominates
the central component. In many of the lower-luminosity
systems in Helsdon \& Ponman's sample, however, the central component is
dominant.

Helsdon \& Ponman (2000) also compared the $\beta$ values of groups
and rich clusters and found a trend for $\beta$ to decrease as the
temperature of the system decreases. A similar trend had previously
been found in samples of poor and rich clusters (e.g. David et al
1990, White 1991,
Bird et al 1995, Mohr \& Evrard 1997, Arnaud \& Evrard 1999). Mohr et
al (1999) reexamined the effect in clusters and found that it
disappears when the surface brightness profiles are properly modeled
using the two-component beta models. This explanation does not
appear to work for poor groups, however, because Helsdon \& Ponman
(2000) used two-component beta models in their study. The lower
$\beta$ values in groups may be an indication 
that non-gravitational heating has
played a more important role in low-mass systems (David et al 1995,
Knight \& Ponman 1997, Horner et al 1999,
 Helsdon \& Ponman 2000). However, as noted
above, simulations indicate that the derived $\beta$ value depends
strongly on the radii over which the surface brightness fit is performed.
Thus, given the strong correlation between system temperature and 
X-ray extent (Figure 3), 
conclusions about how $\beta$ varies with temperature (i.e. mass)
may be premature.

\subsection{Spectral Properties}

X-ray spectral studies of groups have followed the techniques
previously used for other diffuse X-ray sources such as elliptical
galaxies and rich clusters. The observed data from X-ray instruments
such as ROSAT or ASCA do not give the actual spectrum of the source but a
convolution of the source spectrum with the instrument response. In
general, it is not possible to uniquely invert the convolution and
obtain the input spectrum. The usual solution is to adopt a model
spectrum with a few adjustable parameters and to find the best fit to
the observed data. By analogy to rich clusters, it has generally been
assumed that the dominant emission mechanism in groups is thermal
emission from diffuse, low-density gas. Many authors have calculated
the spectrum emitted by a hot, optically thin plasma. The most popular
models are that of Raymond \& Smith (1977) and Mewe and collaborators
(the so-called MEKAL model; Mewe et al 1985, Kaastra \& Mewe 1993,
Liedahl et al 1995). For simplicity, single-temperature
(i.e. isothermal) models are usually assumed. The free parameters of
interest in the isothermal plasma models include the gas temperature
and metal abundance. For very hot systems, such as rich clusters, the
X-ray emission in the isothermal model 
is dominated by the free-free continuum from hydrogen
and helium. For the temperatures more typical of groups ($\sim$ 10$^7$
K), much of the flux is found in line emission and bound-free continuum.

\subsubsection{GAS TEMPERATURE}
In general, isothermal plasma models provide good fits to the
ROSAT PSPC spectra of groups. The derived gas temperatures are in the
range $\sim$ 0.3--1.8 keV (see Figure 3), which is roughly what is
expected given the range of observed velocity dispersions for groups
(e.g. Ponman et al 1996, Mulchaey et al 1996a, Mulchaey \& Zabludoff
1998, Helsdon \& Ponman 2000). 
There is generally good agreement in
the literature on the temperature of the gas; multiple authors
have derived temperature values within 10\% of each other, even when
temperatures were derived over vastly different physical apertures
(e.g. Mulchaey et al 1996a).
The temperatures derived from the different plasma models
(i.e. Raymond-Smith, MekaL) are also fairly consistent with each other
(e.g. Mulchaey \& Zabludoff 1998). Furthermore, there is very good agreement
between gas temperatures determined by the ROSAT PSPC and ASCA for
systems with temperatures less than about 2 keV. For 
higher-temperature gas (i.e. clusters), the ROSAT data appear to
underestimate the true gas temperature by approximately 30\% (Hwang et
al 1999). All these observations suggest that the derived
temperatures for the intragroup medium are fairly robust.

For some of the groups observed by ROSAT, it is possible to measure
temperature profiles for the hot gas (Ponman \& Bertram 1993, David et
al 1994, Doe et al 1995, Davis et al 1996, Trinchieri et al 1997,
Mulchaey \& Zabludoff 1998, Helsdon \& Ponman 2000, Buote 2000c).
These profiles
suggest that the gas is not strictly isothermal, but rather follows a
somewhat universal form: the gas temperature is at a minimum at the
center of the group, rises to a temperature maximum in the inner
$\sim$ 50--75 h$_{\rm 100}^{-1}$ kpc, and drops gradually at large
radii. The temperature minimum in the inner regions of the group is
coincident with the sharp rise in the X-ray surface brightness
profile. This behavior is consistent with that expected from a \lq\lq
cooling flow\rq\rq \ (cf Fabian 1994). The temperature drop at larger
radii is often based on lower-quality spectra, and in most cases is not
statistically significant. Even if this latter effect is present, the
gas temperature at large radii is usually within 10--15\% of the
temperature maximum. Therefore, isothermality is not a bad assumption
over most of the group, as long as the central regions are
excluded. However, when global gas temperatures are quoted for groups
in the literature, the central regions are almost always
included. Because the central regions dominate the total counts in the
spectrum, the temperatures found in the literature may
underestimate the global temperatures in many cases.

\subsubsection{$\beta$$_{\rm spec}$}
Although most authors have estimated the ratio of specific energy of
the galaxies to the specific energy of the gas (i.e. the $\beta$
parameter) from surface brightness profiles (see Section 3.3.3),
$\beta$ can in principle be determined by directly measuring $\sigma$
and T$_{\rm gas}$. Unfortunately, because $\sigma$ is usually derived
from only a few velocity measurements, this method is often not very
robust. Detailed membership studies have been made for a few X-ray
groups (i.e. Ledlow et al 1996, Zabludoff \& Mulchaey 1998, Mahdavi
et al 1999), and in these cases the velocity dispersion estimates are
more reliable. Using such estimates, Mulchaey \& Zabludoff (1998)
found $\beta$$_{\rm spec}$ $\sim$ 1 for most of the groups in their
sample. Helsdon \& Ponman (2000) found a similarly high value for
$\beta$$_{\rm spec}$ for groups with temperatures of $\sim$ 1 keV, but
noted a trend for $\beta$$_{\rm spec}$ to decrease in the 
lower-temperature systems. However, almost all of the low-temperture
groups in the Helsdon \& Ponman (2000) sample have
velocity dispersions determined from a small number of galaxies.
Thus, while the current data suggest 
a trend for $\beta$$_{\rm spec}$ to decrease as the temperature of the 
group decreases, detailed spectroscopy of cool groups will be 
required to verify this result.

The $\beta$ $\sim$ 1 values derived for hot groups from the direct
measurement of temperature and velocity dispersion ($\beta$$_{\rm spec}$)
are significantly
higher than the values of $\beta$ often derived from surface brightness 
profile fits ($\beta$$_{\rm fit}$).
This so-called $\beta$-discrepancy problem has been 
discussed extensively for rich clusters (e.g. Mushotzky 1984, Sarazin 1986,
Edge \& Stewart 1991, Bahcall \& Lubin 1994). Based on simulations,
Navarro et al (1995) concluded that $\beta$$_{\rm fit}$ is 
biased low in galaxy clusters because of the limited radial range used 
in the X-ray profiles. This explanation may also explain the 
discrepancy found for groups, which are typically detected to a much 
smaller fraction of the virial radius than their rich cluster 
counterparts. Therefore, the $\beta$-discrepancy in groups may be an 
indication that the current derived $\beta$$_{\rm fit}$ values underestimate 
the true $\beta$ values in many cases.

\subsubsection{GAS METALLICITY}
In addition to measuring gas temperatures, ROSAT PSPC and ASCA
observations of groups have been used to estimate the metal content of
the intragroup medium. As noted earlier, X-ray spectra of groups
are dominated by emission line features. The strongest emission lines
are produced when an electron in a highly ionized atom is collisional
excited to a higher level and then radiatively decays to a lower
level. The most important features in the X-ray spectra of groups
include the K-shell (n=1) transitions of carbon through sulfur and the
L-shell (n=2) transitions of silicon through iron. Particularly
important is the Fe L-shell complex in the spectral range $\sim$
0.7--2.0 keV (Liedahl et al 1995). The wealth of line features in the
soft X-ray band potentially provides powerful diagnostics of the
physical conditions of the gas including the excitation mechanism
and the elemental abundance (Mewe 1991, Liedahl et al
1990).

Unfortunately, the X-ray telescopes flown to date have not had high
enough spectral resolution to resolve individual line complexes.
Still, many attempts have been made to estimate the elemental
abundance of the gas. For groups, this method primarily
measures the iron abundance in the gas, because lines from this
element dominate the spectra. Spectral fits to both ROSAT and ASCA
data suggest that the metallicity of the intragroup medium varies
significantly from group to group; some systems are very 
metal-poor ($\sim$ 10--20\% solar), whereas others are more enriched ($\sim$
50--60\% solar; Mulchaey et al 1993; Ponman \&
Bertram 1993; David et al 1994; Davis et al 1995; Saracco \& Ciliegi
1995; Davis et al 1996; Ponman et al 1996; Mulchaey et al 1996a;
Fukazawa et al 1996, 1998; Mulchaey \& Zabludoff 1998; Davis et al
1999; Finoguenov \& Ponman 1999; Hwang et al 1999; Helsdon \& Ponman
2000). The low metallicities measured in some groups are surprising
because the ratio of stellar mass to gas mass is higher in groups than
in clusters. Consequently, one would naively expect the metallicities of the
gas to be higher in groups than in rich clusters.

Several potential problems have been noted with the
low metallicity measurements for
the intragroup medium.
Ishimaru \& Arimoto (1997) pointed out that
most X-ray studies have adopted the old photospheric value for the solar
Fe abundance (Fe/H $\sim$ 4.68 $\times$
10$^{-5}$), whereas the commonly accepted
\lq\lq meteoritic\rq\rq \ value
is significantly lower (Fe/H $\sim$ 3.24 $\times$ 10$^{-5}$). (Note
that more recent estimates of the photospheric Fe abundance in the sun
are consistent with the meteoritic value; see McWilliam 1997).
Thus, essentially all the Fe measurements in the X-ray literature
should be increased by a factor of 
$\sim$ 1.44 to renormalize to the meteoritic value. This is particularly
important when comparing the X-ray metallicities to 
chemical-evolution models, which usually adopt the 
meteoritic Fe solar abundance. 
The ability of ROSAT
data to properly measure the gas abundance has also been questioned. Bauer
\& Bregman (1996) measured metallicities with the ROSAT PSPC for stars
with known metallicities close to the solar value, and found the ROSAT
metallicities were typically a factor of five lower than the optical
measurements. Bauer \& Bregman (1996) suggested several possible
explanations for the discrepancy, including instrumental calibration
uncertainties, problems with the plasma codes and possible differences
in the photospheric and coronal abundances of stars. Instrumental
uncertainties with the ROSAT PSPC are unlikely to be the major source
of the problem because ASCA spectroscopy of groups also indicates low
gas metallicities (Fukazawa et al 1996, 1998; Davis et al 1999;
Finoguenov \& Ponman 1999; Hwang et al 1999). The possibility that the
plasma models are inaccurate or incomplete has been a major
concern. While abundance measurements for rich clusters are derived
primarily from the well-understood Fe K-$\alpha$ line, group
measurements rely on the much more complicated Fe L-shell
physics. Problems with the plasma models were in fact identified by
early ASCA observations of cooling flow clusters (Fabian et al
1994). Liedahl et al's (1995) revision to the standard MEKA thermal
emission model likely accounts for the largest problems in the earlier
plasma codes. However, fits to ASCA spectra of groups with the revised
model still require very low metal abundances. Hwang et
al (1997) have shown that for clusters with sufficient Fe L and Fe K
emission (i.e. clusters with temperatures in the range $\sim$ 2--4
keV), the metallicities derived from the Fe L line complex are
consistent with the values derived from the better understood Fe K
complex (see also Arimoto et al's 1997 analysis of the Virgo cluster).
Unfortunately, it is not clear that the reliability of the Fe L diagnostics 
implied from $\sim$ 2--4 keV poor clusters
necessarily extends down to lower temperature groups,
since other Fe lines 
dominate the spectrum below $\sim$ 1 keV (Arimoto et al 1997).
Therefore, some problems with the plasma models may still
exist. 

Another potentially important problem is that the usually assumed
 isothermal model may be inappropriate for groups (Trinchieri et al
 1997; Buote 1999,2000a). There is clear evidence for temperature
 gradients in groups, particularly in the inner $\sim$ 50 h$_{\rm
 100}$$^{-1}$ kpc. In fact, the surface brightness profiles of ROSAT PSPC data
 suggest the presence of at least two distinct components in groups
 (Mulchaey \& Zabludoff 1998). Mixing of multiple-temperature
 components is particularly an issue for ASCA data because separating
 out the central component from more extended emission is not possible
 with the ASCA point spread function. Buote (1999,2000a) has studied this
 problem in detail for both elliptical galaxies and groups, and finds
 that in general single-temperature models provide poor fits to the
 ASCA spectra. By adopting a two-temperature model,
one can obtain better fits, 
 and the metallicities derived are substantially higher. For
 a sample of 12 groups, Buote (2000a) derives an average metallicity of Z =
 0.29$\pm{0.12}$ Z$_{\odot}$ for the isothermal model and Z =
 0.75$\pm{0.24}$ Z$_{\odot}$ for the two-temperature model (a single
 metallicity is assumed for the gas in these models). Buote (2000a)
 also finds that a multiphase cooling flow model provides a good
 description of the data. This model also requires higher
 metallicities (Z=0.65$\pm{0.17}$ Z$_{\odot}$). Buote (2000a) finds a trend
 for the metallicities to be lowest in those groups for which the
 largest extraction apertures were used. This result is consistent
 with metallicity gradients in groups (see also Buote 2000c).
 Alternatively, it may simply
 reflect that the relative contribution of the \lq\lq group\rq\rq \
 gas component increases as one adopts a larger aperture. In fact,
 given the results of the ROSAT surface brightness profile fits,
emission from the central elliptical galaxy
may dominate the flux in the
 typical ASCA aperture and thus likely dominates the metallicity
 measurement. Therefore, the ASCA measurements may not be providing an
 accurate gauge of the global metal content of the group gas.
Regardless, the work of Buote (1999,2000a) is an important reminder
that 
the properties derived from X-ray spectroscopy are very sensitive to the
choice of the input model.

Matsushita et al (2000)
also considered multi-temperature models 
for a large sample of early-type galaxies observed with ASCA. In contrast to
Buote (1999, 2000a),
Matsushita et al (2000) 
concluded that the poor spectral fits to ASCA data
were not caused by incorrect
modeling of multi-temperature emission. 
Furthermore, the multi-temperature models used by Matsushita et al (2000)
produced relatively small increases in the overall abundance in many cases.
Matsushita et al (2000) suggested 
that the strong coupling between the abundance of the so-called 
$\alpha$-elements (i.e. O, Ne, Mg, Si, S) and the abundance of Fe 
hampers a unique determination of the overall metallicity. By fixing the 
abundance of the $\alpha$-elements, Matsushita et al (2000) found that the 
derived metallicities are approximately solar.
Although Matsushita et al
(2000) restricted their analysis to early-type galaxies, these results
may be applicable to groups, which have X-ray properties very similar 
to those of X-ray luminous ellipticals.

Although the dominant line features for the intragroup medium are
produced by iron, strong lines are also expected from elements such as
oxygen, neon, magnesium, silicon, and sulfur. The relative abundance of
these various elements provides strong constraints on the star
formation history of the gas. Some authors have attempted to fit the
ASCA spectra with an isothermal model where the 
$\alpha$-elements 
are varied together and separately from the iron abundance
(Fukazawa et al 1996, 1998; Davis et al 1999;
Finoguenov \& Ponman 1999; Hwang et al 1999). In general, these
studies find that the $\alpha$-element to iron ratio is approximately
solar in groups. Unfortunately, the determination of this ratio is
very sensitive to the spectral model adopted (Buote 2000a) and if the
isothermal assumption is not valid, these determinations are not
particularly meaningful. 

In summary, despite the great potential of X-ray spectroscopy to provide
clues into the enrichment history of the intragroup medium, it is not
possible at the present time to make strong conclusions about the metal
content of the hot gas. Until we have higher resolution X-ray
spectra and more complete plasma codes, the metallicity of the
intragroup medium will remain an open issue.

\subsubsection{ABSORBING COLUMN}
The soft X-ray band is sensitive to low-energy photoabsorption by gas
both within the source and along the line of sight. This absorption
must be included in the X-ray spectral fits. It is usually assumed that
the X-ray flux is diminished by:

\centerline{A(E)= exp(-N$_{\rm H}$$\sigma$(E))}

\noindent{where N$_{\rm H}$ is the hydrogen
 column density and $\sigma$(E) is the 
photo-electric cross-section (solar abundances are almost universally 
assumed for the 
absorbing gas). The cross sections in Morrison \& McCammon (1983) are 
commonly adopted for X-ray analysis. The standard procedure is to 
allow N$_{\rm H}$ to be a free parameter in the spectral fit. If the 
best-fit spectral model returns a value of N$_{\rm H}$ significantly
higher than the Galactic value, this is taken as evidence for excess 
absorption intrinsic to the group or central galaxy. 
The ROSAT and ASCA spectra of groups are often not of high enough quality
to adequately constrain the absorbing column. Therefore, 
many authors have chosen to fix N$_{\rm H}$ to the Galactic value
for spectral fits. For a few groups,
however, column densities above the Galactic value have been inferred
(Fukazawa et al 1996; Davis et al 1999; Buote 2000a,b). Buote (2000b) 
undertook the most ambitious study of absorption in groups, measuring
N$_{\rm H}$ as a function of radius in a sample of 10 luminous systems 
observed by the ROSAT PSPC. Buote (2000b) found that the value of
N$_{\rm H}$ derived depends strongly on the bandpass used in the X-ray
analysis and suggested the bandpass-dependent 
N$_{\rm H}$ values are consistent with
additional absorption in
the group from a collisionally ionized gas. This excess absorption
manifests itself primarily as a strong oxygen edge feature at 
$\sim$ 0.5 keV. Buote (2000b) found that within the central
regions of the groups, the estimated masses of the absorbers are 
consistent with the matter deposited by a cooling flow over the 
lifetime of the flow. 
If a warm absorber exists in groups, as suggested by Buote (2000b),
it should be verified by the 
next generation of X-ray telescopes.}

\subsubsection{X-RAY LUMINOSITY}
For a thermal plasma, the X-ray luminosity is a rough measure of the
total mass in gas. Therefore, the total X-ray luminosity of a group
provides a potentially interesting probe of a group's properties. In
almost all cases in the literature, the total flux or luminosity
quoted is out to the radius to which X-ray emission is detected. In
this sense, quoted X-ray luminosities should be thought of as \lq\lq
isophotal luminosities\rq\rq\ . 
The measured luminosity is
also sensitive to the exact techniques used in the X-ray analysis. For
example, the total radial extent of the X-ray emission (and thus the
total X-ray luminosity) is strongly dependent on the assumed
background level (Henriksen \& Mamon 1994, Davis et al 1996). Because
of this, different authors often derive vastly different X-ray
luminosities for the same group using the same ROSAT observation
(Mulchaey et al 1996a).

It is a common practice to quote bolometric luminosities in the
literature. The bolometric correction is estimated by extrapolating
the spectral model for the gas beyond the limited bandpass of the
particular telescope and by making a correction for any absorption
along the line- of- sight. In the case of ROSAT observations, these
corrections can easily double the luminosity of the source. The
bolometric correction is also somewhat sensitive to uncertainties in
the spectral model such as gas metallicity. For very shallow
observations, such as those based on ROSAT All-Sky Survey data, a spectral
model must usually be assumed to estimate the total X-ray luminosity.  The
bolometric luminosities of groups are typically in the range several
times 10$^{40}$ h$_{\rm 100}$$^{-2}$ to nearly 10$^{43}$ h$_{\rm 100}$$^{-2}$
(Mulchaey et al 1996a, Ponman et al 1996, Helsdon \& Ponman 2000).
Thus, the X-ray luminosities of groups can be several orders of
magnitude lower than the X-ray luminosities of rich clusters
(cf Forman \& Jones 1982).

Finally, it is worth noting that because X-ray emission is usually traced
only to a fraction of the virial radius in groups, it is likely that
the isophotal measurements significantly underestimate the true
luminosities of the hot gas.
This is particularly true for the coolest groups. 
Helsdon \& Ponman (2000) have attempted
to account for the missing luminosity by extrapolating the gas density
profile models out to the virial radius. A comparison of the observed
isophotal luminosities to the corrected virial luminosities in the
Helsdon \& Ponman sample indicate that in many cases, over half of the
luminosity could occur beyond the radius to which X-ray emission is
currently detected.

\section{CORRELATIONS}

There has been considerable interest in how the X-ray and optical
properties of groups differ from those of richer clusters. Such
comparisons are often limited by the poorly determined group
properties. Most optical properties of groups are derived from
existing redshift surveys, which typically only include the most
luminous group members. Consequently, global properties such as
velocity dispersion and morphological composition are subject to small
number uncertainties. The properties of the hot gas also tend to be
more uncertain in poorer systems than in clusters because of the lower
X-ray fluxes of groups. 
It should also be
remembered that the X-ray properties of groups and clusters are often
derived over very different gas density contrasts, which further complicates
the comparisons of these systems. Despite
these potential problems, group and cluster comparisons have
provided considerable insight into the nature of X-ray groups.

\subsection{T-$\sigma$ Relation}
Because both the temperature of the intragroup medium and the velocity
dispersion of the galaxies provide a measure of the gravitational
potential strength, a correlation between these two quantities is
expected. Although there is considerable scatter in the data, ROSAT
observations are consistent with such a correlation (Figure 4; Ponman et al
1996, Mulchaey \& Zabludoff 1998, Helsdon \& Ponman 2000). 
High-temperature groups (T $\sim$ 1 keV) appear to follow the extrapolation
of the trend found for rich clusters; the ratio of specific energy
in the galaxies to specific energy in the gas is approximately one
(i.e. $\beta$ $\sim$ 1 and T $\propto$ $\sigma$$^2$; Mulchaey \&
Zabludoff 1998, Helsdon \& Ponman 2000). 
Ponman et al (1996) and Helsdon \& Ponman (2000) 
have claimed that the T-$\sigma$ relation becomes much
steeper for cooler groups. However, Figure 4 suggests that given
the large scatter, evidence for a systematic
deviation from the cluster relationship is at 
this point rather scarce. 

\begin{figure}
\psfig{figure=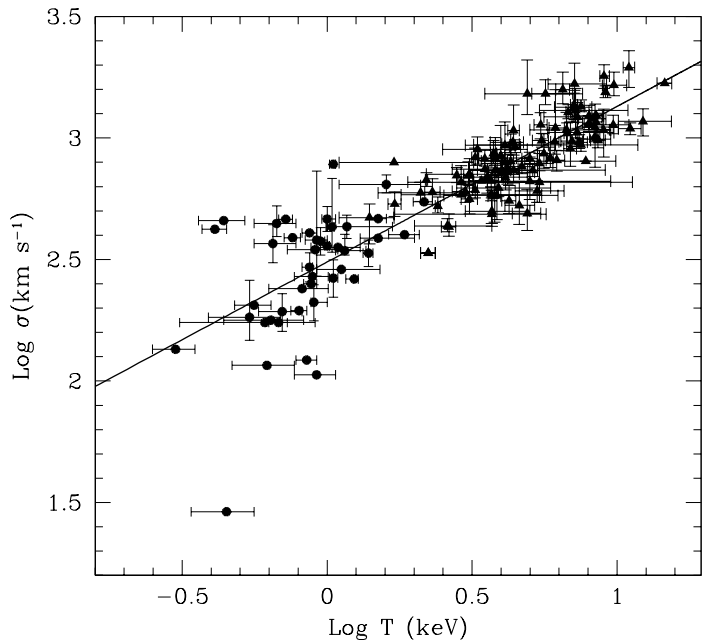,width=4.0in}
\caption{Logarithm of the X-ray temperature versus logarithm of optical
velocity dispersion for a sample of groups (circles) and clusters (triangles).
The group data are taken from the literature compilation of Xue \& Wu (2000),
with the addition of the groups in Helsdon \& Ponman (2000). The cluster
data are taken from Wu et al (1999). The solid line represents the 
best-fit found by Wu et al (1999) for the clusters sample (using an 
orthogonal distance regression method). Within the large scatter, the groups
are consistent with the cluster relationship.} 
\end{figure}

\begin{figure}
\psfig{figure=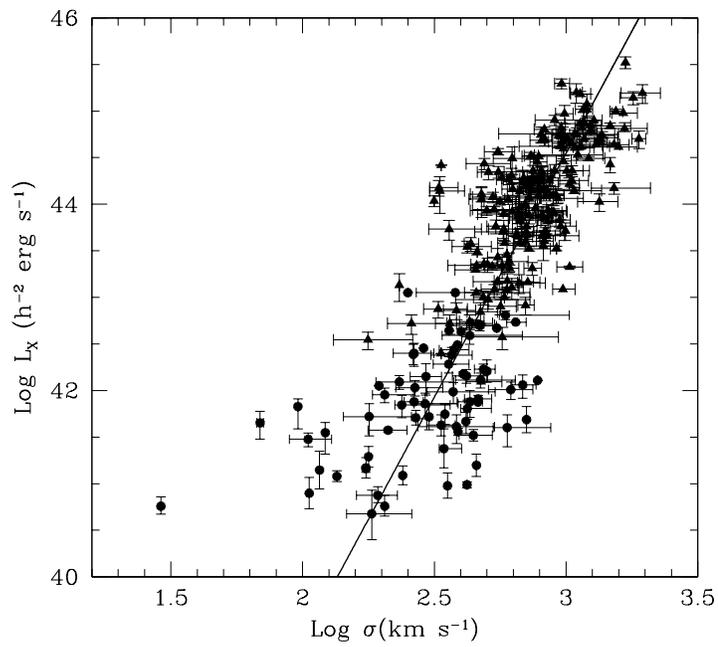,width=4.0in}
\caption{Logarithm of optical
velocity dispersion versus logarithm of X-ray luminosity
 for a sample of groups (circles) and clusters (triangles). The data are
taken from the same sources cited in Figure 4.
The solid line represents the 
best-fit found by Wu et al (1999) for the clusters sample (using an 
orthogonal distance regression method).}
\end{figure}

\begin{figure}
\psfig{figure=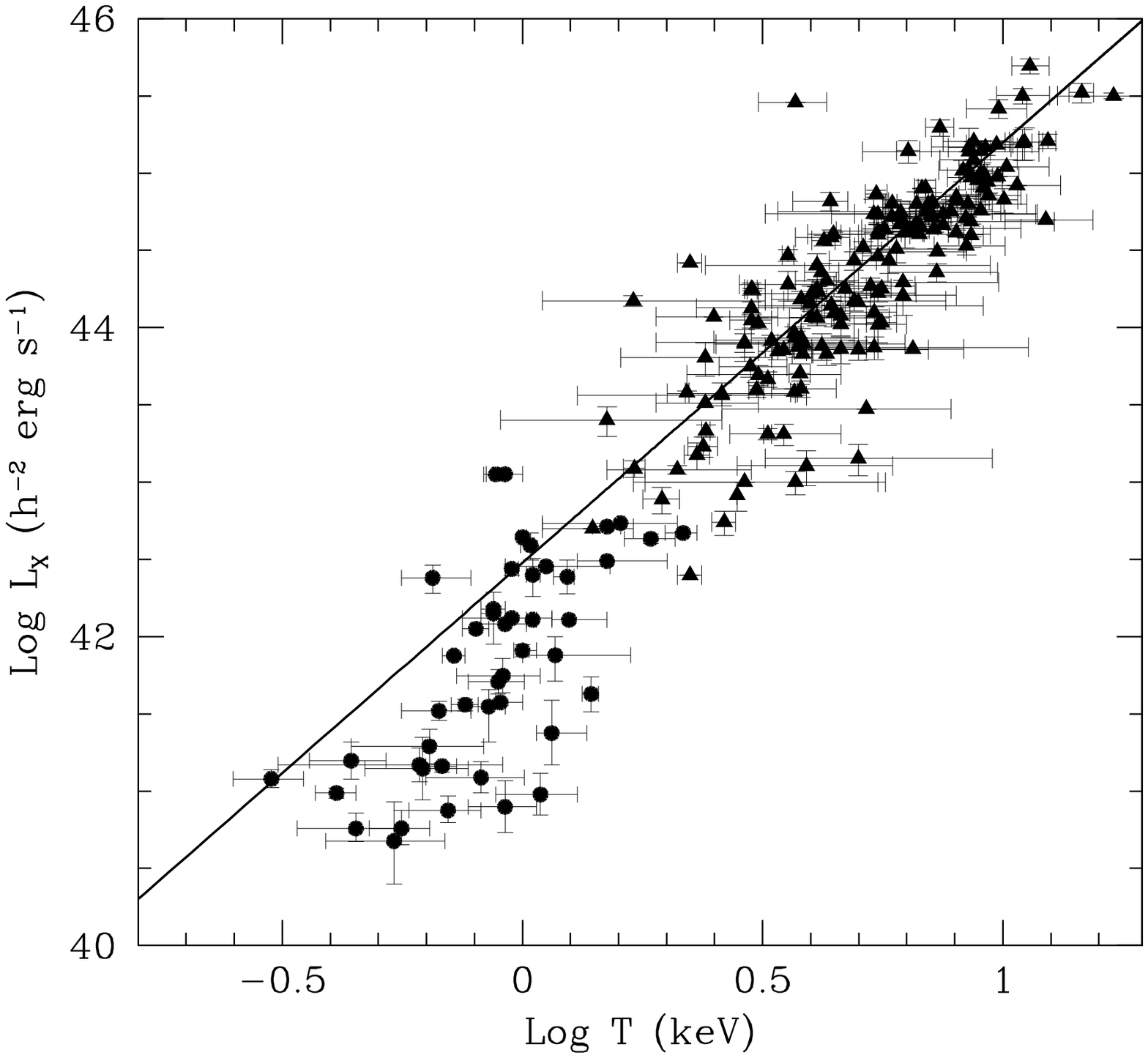,width=4.0in}
\caption{Logarithm of the X-ray temperature
versus logarithm of X-ray luminosity
 for a sample of groups (circles) and clusters (triangles). The data are
taken from the same sources cited in Figure 4.
The solid line represents the 
best-fit found by Wu et al (1999) for the clusters sample (using an 
orthogonal distance regression method). The observed relationship for
groups is somewhat steeper than the best-fit cluster relationship.}
\end{figure}

\subsection{L$_{\rm X}$-$\sigma$ 
and L$_{\rm X}$-T Relations}
Strong correlations are also found between X-ray luminosity and both
velocity dispersion and gas temperature in groups.
However, there is
considerable disagreement in the 
literature over the nature of these correlations.
Figure 5 shows the L$_{\rm X}$-$\sigma$ relationship for all the groups
observed by the ROSAT PSPC in pointed-mode and a sample of clusters
observed with various X-ray telescopes (Wu et al 1999). The solid line
shows the best-fit relationship Wu et al (1999) derived from the cluster
sample alone. Figure 5 shows that for the most part, groups are 
consistent with the cluster relationship, although there is considerable
scatter particularly among the lowest luminosity groups. This conclusion
was reached by 
Mulchaey \& Zabludoff (1998), who found that a single relationship fit 
their sample of groups and rich clusters. 
Ponman et al (1996) and Helsdon \& Ponman (2000) also found that 
the L$_{\rm X}$-$\sigma$ for groups was basically consistent with 
the cluster relationship, although both studies noted that the relationship
may become somewhat flatter for low velocity dispersion systems. (Within the 
errors, the slopes derived by Mulchaey \& Zabludoff, Ponman et al (1996)
and Helsdon \& Ponman (2000) are indistinguishable;
L$_{\rm X}$ $\propto$ $\sigma$$^{4.3}$, $\sigma$$^{4.9}$ and $\sigma$$^{4.5}$,
respectively). Therefore, there is fairly good agreement among the ROSAT studies
based on pointed-mode data.
However, Mahdavi et al (1997) derived a significantly
flatter slope from their ROSAT All Sky Survey data (L$_{\rm X}$
$\propto$ $\sigma$$^{1.56}$) and suggested that for low velocity dispersion
systems the X-ray emission is dominated by hot gas clumped around 
individual galaxies. More recently, Mahdavi et al (2000) presented 
X-ray luminosities for a much larger sample of loose groups. In agreement
with their earlier result, they find a much flatter L$_{\rm X}$-$\sigma$ 
for groups than for rich clusters. Mahdavi et al (2000) modeled the 
L$_{\rm X}$-$\sigma$ relationship as a broken power law, with a very flat
slope (L$_{\rm X}$ $\propto$ $\sigma$$^{0.37}$) for systems with velocity
dispersion less than 340 km s$^{-1}$ and a cluster-like value 
(L$_{\rm X}$ $\propto$ $\sigma$$^{4.0}$) for higher velocity dispersion
systems. However, a visual inspection of 
Mahdavi et al's (2000)
L$_{\rm X}$-$\sigma$ relationship
(see Figure 4 of their paper) reveals that the need for a broken power law
fit is driven by the one or two lowest velocity dispersion groups (out of
a total sample of 61 detected groups.) Furthermore, nearly all the
L$_{\rm X}$ upper
limits derived by 
Mahdavi et al (2000) fall below their broken power law relationship 
(and therefore require a \lq\lq steeper\rq\rq \ relationship).
Thus, the case for 
deviations from the L$_{\rm X}$-$\sigma$ cluster relationship is far 
from compelling. It is also worth noting that the velocity 
dispersions of the  groups that appear to 
deviate the most from the cluster relationship are often based on very
few velocity measurements (for example the most \lq\lq deviant\rq\rq \
system in 
Figures 4 and 5 
 has a velocity dispersion based on only four velocity measurements.)
Zabludoff \& Mulchaey (1998) have found that when velocity dispersions
are calculated for X-ray groups from a large number of galaxies, as opposed
to just the four or five brightest galaxies, the velocity dispersion is 
often significantly underestimated. Therefore, more detailed velocity 
studies of low velocity dispersion groups could prove valuable in verifying 
deviations from the cluster L$_{\rm X}$-$\sigma$ relation.

There is also considerable disagreement in the literature about the
relationship between X-ray luminosity and gas temperature. Mulchaey
\& Zabludoff (1998) found that a single L$_{\rm X}$--T relationship
could describe groups and clusters
(L$_{\rm X}$ $\propto$
T$^{2.8}$). 
However, both Ponman et al (1996) and Helsdon \& Ponman (2000) found
much steeper relationships for groups (L$_{\rm X}$ $\propto$ T$^{8.2}$
and L$_{\rm X}$ $\propto$ T$^{4.9}$, respectively). These differences
might be attributed to the different temperature ranges included in
the studies. Mulchaey \& Zabludoff's (1998) sample was largely
restricted to hot groups (i.e. $\sim$ 1 keV), whereas Ponman and
collaborators have included much cooler systems (down to $\sim$ 0.3
keV). Indeed, Helsdon \& Ponman (2000) found that the steepening of
the L$_{\rm X}$-T relationship appears to occur below about 1 keV.
Figure 6 suggests that the deviation of the cool groups from the 
cluster relationship is indeed significant.
The fact that the L$_{\rm X}$-$\sigma$ relationship for groups appears
to be similar to the relationship found for clusters, while the
relationships involving gas temperature significantly depart from the
cluster trends, may be an indication that 
non-gravitational heating is important in groups (Ponman et al 1996,
Helsdon \& Ponman 2000). However, 
the group X-ray luminosities may be biased somewhat low because groups
are detected to a smaller fraction of their virial radius than richer
systems and if comparisons are made at the same mass over-density level,
groups would likely fall closer to the cluster relation.

\subsection{Galaxy Richness and Optical Luminosity}
Most authors have found little or no correlation between X-ray
luminosity and the number of luminous galaxies in a group or the total
optical luminosity of the group (Ebeling et al 1994, Doe et al 1995,
Mulchaey et al 1996a, Ponman et al 1996).  The lack of correlation
between X-ray luminosity and number of group members is not too
surprising because galaxy-galaxy merging is likely prevalent in groups,
and thus the number of galaxies in a group is likely not conserved in
time (Ponman et al 1996). The fact that there is no relationship
between optical and X-ray luminosity is important because it suggests
that the X-ray emission is not associated with individual galaxies for
most of the samples studied (Ponman et al 1996).
 
Mahdavi et al (1997) came to a very different conclusion with their
RASS survey of optically-selected groups: They found a strong correlation
between X-ray luminosity and optical luminosity. The differences
between Mahdavi et al's (1997) results and those of other authors
suggests that the groups in Mahdavi et al (1997) may be systems
dominated by X-ray emission from individual galaxies and not
intragroup gas.

\subsection{Morphological Content} Correlations between the
presence of X-ray emission and the morphological composition of groups
were suggested from the earliest ROSAT studies. Ebeling et al (1994)
were the first to claim such an effect, noting that all but one of the
X-ray detected HCGs in the ROSAT All-Sky Survey data had spiral
fraction less than 50\%. Subsequent studies of small samples appeared
to support this trend (Henry et al 1995, Pildis et al 1995, Mulchaey
et al 1996a). However, Ponman et al (1996) came to a very different
conclusion based on their much larger survey of the HCGs. They
detected several groups with high spiral fractions, including the
extreme example HCG 16, a compact group that contains only spirals.

Figure 7 shows the distribution of early-type fraction for all the
groups with published pointed observations with ROSAT. For the
purposes of this plot, a group is considered \lq\lq X-ray
detected\rq\rq \ only if there is evidence for an extended intragroup
medium component. As is apparent from this figure, a significant
number of spiral-rich groups do contain diffuse X-ray emission,
which confirms the conclusion of Ponman et al (1996). In fact, in contrast
to the earlier studies, the distribution of early-type fractions is
surprisingly flat for the X-ray detected groups. The apparent
contradiction with the earlier results can be explained by the fact
that the majority of the groups in the current sample were selected
from optical redshift surveys and were serendipitously observed by
ROSAT (Helsdon \& Ponman 2000, Mulchaey et al 2000), whereas the earlier
studies were biased toward X-ray luminous groups, which tend to have
higher early-type fractions (Mulchaey \& Zabludoff 1998).

\begin{figure}
\psfig{figure=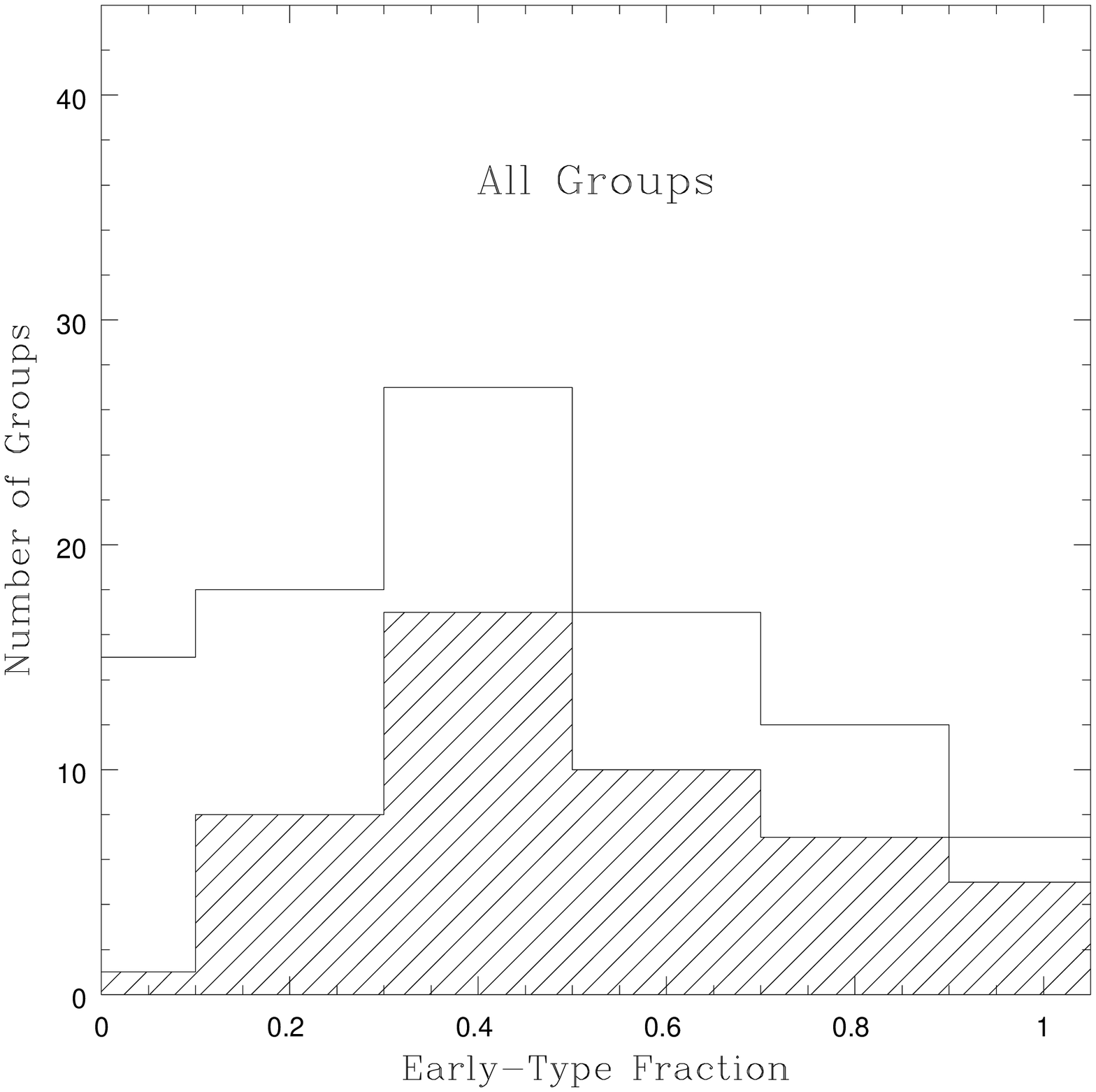,width=4.0in}
\psfig{figure=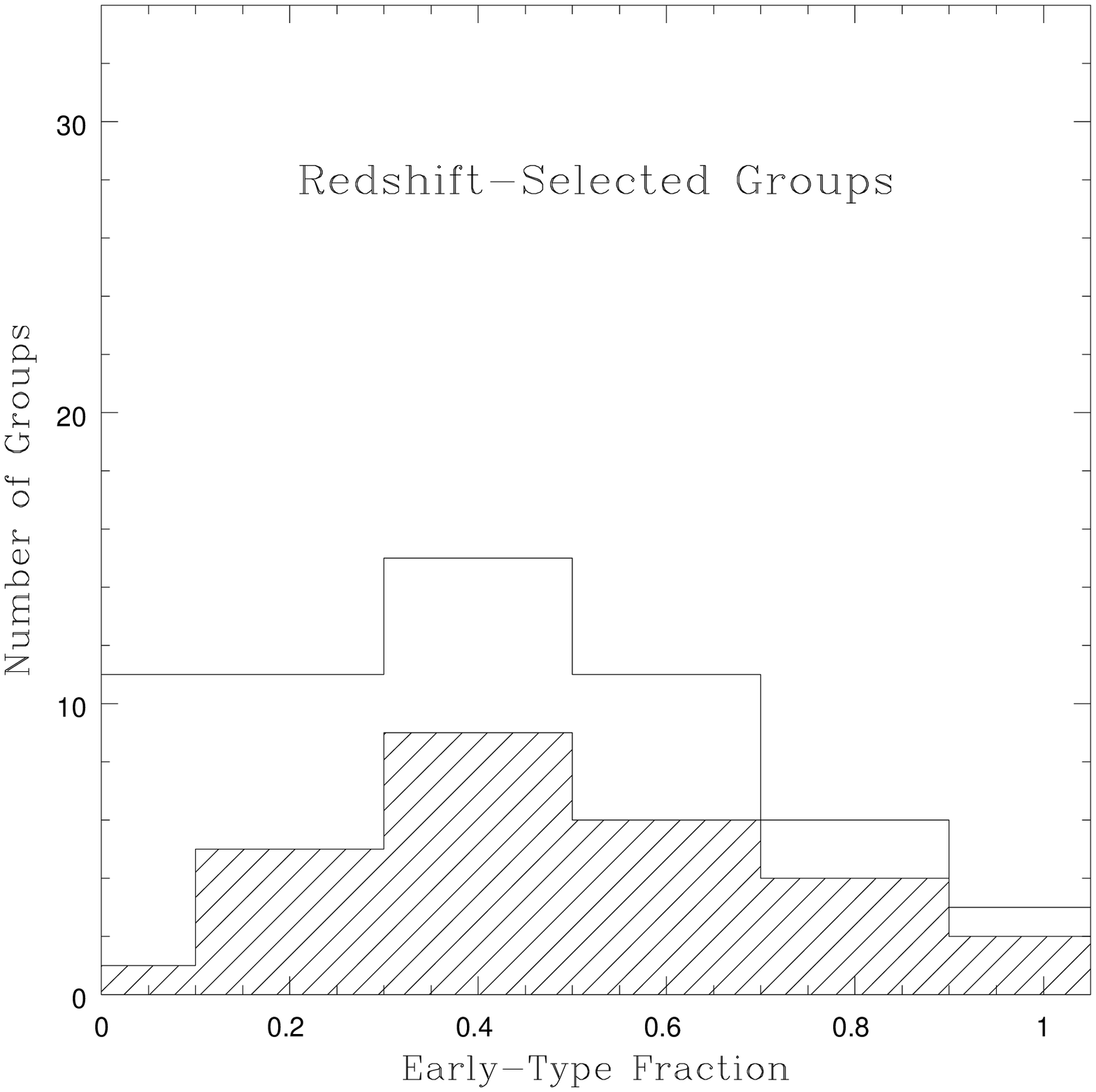,width=4.0in}
\caption{Distribution of early-type fraction for all groups ({\it open histogram})
and groups with diffuse X-ray emission ({\it shaded histogram}). The top panel
shows the result for all published PSPC pointed-mode observations, whereas the
lower panel contains only groups selected from optical redshift surveys.} 
\end{figure}

A closer examination of Figure 7 reveals that while many spiral-rich
systems are X-ray sources, spiral-only groups tend not to contain a
diffuse X-ray component. The one exception in Figure 7 is HCG 16.
However, the true nature of the X-ray emission in HCG 16 is unclear.
The ROSAT image of the group indicates that the emission is very clumpy and
concentrates around the brightest group members (see Figure 2). Some
authors have attributed all of the X-ray emission to individual
galaxies (Saracco \& Ciliegi 1995; see also an earlier Einstein
observation by Bahcall et al 1984), whereas others have claimed the
existence of intragroup gas (Ponman et al 1996). Dos Santos \& Mamon
(1999) have reanalyzed the ROSAT PSPC data on HCG 16, paying special
attention to the removal of emission associated with galaxies.
Although Dos Santos \& Mamon (1999) derived a lower luminosity for the
diffuse gas than Ponman et al (1996), they still found evidence for
some diffuse gas. However, the presence of diffuse emission does not
necessarily mean that HCG 16 contains a diffuse intragroup medium. One
possibility is that the emission is related to the unusually high
number of active galaxies in the group (HCG 16 contains one Seyfert
galaxy, two LINERs, and three starburst galaxies; Ribeiro et al 1996).
The X-ray to infrared luminosity ratio of this system is much higher
than one would expect if the X-ray emission is related to the
galaxies' activity, however (Ponman, private
communication). Alternatively, the X-ray emission may be associated
with shocked gas, as appears to be the case in Stephan's Quintet
(Pietsch et al 1997).

With the possible exception of HCG 16, all X-ray detected groups
studied to date contain at least one early-type galaxy. There are
several possible explanations for why spiral-only groups do not
contain diffuse X-ray emission. One possibility is that all
spiral-only groups are chance superpositions and not real, physical
systems.  
This possibility seems unlikely, given the existence of our own
spiral-only Local Group (see Section 5.10 for a discussion of the 
intragroup medium in the Local Group).
Another possibility is that the intragroup gas in
spiral-only groups is too cool to produce appreciable amounts of X-ray
emission (Mulchaey et al 1996b).
 Based on velocity dispersions, the virial temperatures of
spiral-only groups do tend to be lower than those of their early-type
dominated counterparts (Mulchaey et al 1996b). While a cool
(i.e. several million degrees K) intragroup medium would be difficult
to detect in X-ray emission, such gas might produce prominent absorption
features in the far-ultraviolet or X-ray spectra of background quasars
(Mulchaey et al 1996b, Perna \& Loeb 1998, Hellsten et al 1998).
In fact, several such groups may have already
been detected as OVI $\lambda$$\lambda$1031.93,1037.62 \AA \
absorption systems (Bergeron et al 1994, Savage et al 1998). A third
possibility is that the gas densities in spiral-only groups are too
low to be detected in X-rays. Low gas densities in spiral-only groups
are in fact consistent with recent prediction of preheating models
for groups
(Ponman et al 1999; see Section 5.9).

\section{COSMOLOGICAL IMPLICATIONS OF X-RAY GROUPS}

\subsection{The Physical Nature of Groups}

Simulations of local large-scale structure suggest that a significant
fraction of the groups identified in redshift surveys are not real, bound
systems (Frederic 1995, Ramella et al 1997).
The existence of diffuse X-ray emitting gas is often cited as evidence
that a group is real. This is not necessarily the case, however.
Hernquist et al (1995) noted that primordial gas may be shock-heated to
X-ray emitting temperatures along filaments. When these filaments are
viewed edge-on, a \lq\lq fake\rq\rq \ group with an X-ray halo could
be observed. Ostriker et al (1995) proposed a test of the Hernquist
et al (1995) filament model by defining an observable quantity Q, that
is proportional to the axis ratio of the group. Applying this test to
the early ROSAT observations of HCGs, Ostriker et al (1995) found that
the Q values for most HCGs are consistent with them being
frauds. However, the low Q values for groups can also be explained if
the ratio of gas mass to total mass is smaller in groups than in rich
clusters. Both ROSAT observations (David et al 1995, Pildis et al
1995, Mulchaey et al 1996a) and simulations (Diaferio et al 1994,
Pildis et al 1996) of X-ray groups are in fact consistent with this
idea, suggesting that the Ostriker et al test may in the end not be very
useful.

Several arguments support the idea that at least some X-ray groups are
real, bound systems and that the X-ray gas is virialized. In the most
X-ray luminous groups, the diffuse gas extends on scales of hundreds
of kiloparsecs and appears smooth. This is consistent with what one
expects for a \lq\lq smooth\rq\rq\ group potential. The gas
temperature in these cases agrees fairly well with the temperature
expected based on the velocity dispersion of the groups. Furthermore,
most of these groups show evidence for cooling flows in their centers,
suggesting that the gas is in an equilibrium state and has probably existed
for at least several gigayears.

Ironically, perhaps the best evidence for the reality of the X-ray
luminous groups has come from optical studies of these systems.
Zabludoff \& Mulchaey (1998) used multifiber spectroscopy to study
the faint galaxy population in a small sample of groups and found
large differences in the number of faint galaxies in X-ray detected
and non-detected groups. All of the X-ray detected groups in the
Zabludoff \& Mulchaey (1998) sample contain at least 20--50 group
members (down to magnitudes as faint as M$_{\rm B}$ $\sim$ -14 +5
log$_{\rm 10}$ h$_{\rm 100}$). Even down to these relatively faint
magnitude limits, many of the X-ray detected groups have very high
early-type fractions (nearly 60\% in some cases). The large number of
group galaxies argue that these X-ray groups must be real, physical
systems and not radial superpositions. There are also strong
correlations between dynamical measures of the gravitational potential
(i.e. velocity dispersion/gas temperature) and the early-type fraction
of the group (Zabludoff \& Mulchaey 1998, Mulchaey et al 1998).  These
correlations imply either that galaxy morphology is set by the local
potential at the time of galaxy formation (Hickson et al 1988) or that
the potential grows as the group evolves (Diaferio et al 1993). Either
scenario requires that most X-ray luminous groups be real, bound
systems.

However, it is likely that some \lq\lq X-ray detected\rq\rq \ groups
are not virialized systems.  In particular, low-luminosity, 
low-temperature groups tend to have irregular X-ray morphologies with the
X-ray emission distributed in the immediate vicinity of individual
galaxies. These X-ray morphologies suggest that these groups are
still dynamically evolving. In some cases, such as HCG 92, gas has
apparently reached X-ray emitting temperatures by other mechanisms
such as shocks. Therefore, X-ray detection alone does not indicate
that a
system is virialized.

\subsection{Mass Estimates}
 
One of the most important applications of X-ray observations of groups
has been mass estimates. Prior to ROSAT, mass determinations for
groups were largely based on application of the virial theorem to the
group galaxies. For a typical cataloged group with only four or five
velocity measurements, the virial method can be unreliable
(e.g. Barnes 1985, Diaferio et al 1993).

The method used to estimate group masses from X-ray data is analogous
to the technique developed for rich clusters (e.g. Fabricant et al
1980, 1984; Fabricant \& Gorenstein 1983; Cowie et al 1987). The
fundamental assumption is that the hot gas is trapped in the potential
well of the group and is in rough hydrostatic equilibrium. This
assumption is probably a reasonable one for most groups, given the short
sound crossing times in these systems. A further assumption is that
the only source of heating for the gas is gravitational, i.e. that the
gas temperature is a direct measure of the potential depth and 
therefore of the total mass. This
assumption may not be strictly true for some groups. In particular, the fact
that the heavy metal abundance of the intragroup medium is non-zero
suggests that some of the gas has been reprocessed in the stars in 
galaxies and
ejected by supernovae-driven winds. In addition to polluting the
intragroup gas with metals, such winds also provide additional energy
to the gas. It has generally been assumed in the literature that the
energy contribution of such winds is negligible. Semi-analytic models
suggest that this assumption is fair as long as the temperature of the
system is greater than about 0.8 keV (Balogh et al 1999, Cavaliere et
al 1999). Thus, for many groups, the hydrostatic mass estimator should
be valid.

With the further assumption of spherical symmetry, the mass interior
to radius R is given by (Fabricant et al 1984):

\centerline{M$_{\rm total}$ ($<$R) = ${kT_{\rm
gas}(R)}\over{G{\mu}m_p}$ [${{dlog{\rho}}\over{dlogr}} +
{{dlogT}\over{dlogr}}$] R}
 
\noindent{where k is Boltzmann's constant, T$_{\rm gas}$(R) is the 
gas temperature at radius R, G is the gravitational constant,
 $\mu$ is the mean molecular weight, m$_{\rm p}$ is
the mass of the proton, and $\rho$ is the gas density.
 In principle, all of the unknowns in this
equation can be calculated from the X-ray data. 
Typically, the gas temperature is measured directly from the X-ray spectrum 
and the gas density profile is determined by fitting the standard 
beta model to the surface brightness profile.
Unfortunately, it is
often necessary to make a further assumption that the gas is
isothermal (i.e. ${{dlogT}\over{dlogr}}$= 0). For a few groups,
the temperature profile can be directly measured. The resulting 
mass estimates suggest that the isothermal assumption generally results in an
error in the mass of no more than about 10\% (e.g. David et al 1994, 
Davis et al 1996). With the isothermal assumption, M$_{\rm total}$ ($<$R)
$\propto$ T$_{\rm gas}$$\beta$R (as long as R is much larger than
the core radius in the beta model. Therefore, if $\beta$ is underestimated 
from the surface brightness profile fits by a factor of $\sim$2
(see Sections 3.3.3 and 3.4.3), then the 
mass estimates are also too small by a factor of $\sim$ 2.) 
}

ROSAT measurements indicated a small range of total group
masses with nearly all of the systems clustered around 10$^{13}$
h$^{-1}$ M$_{\odot}$ 
(see Figure 8; Mulchaey et al 1993, Ponman \& Bertram 1993,
David et al 1994, Pildis et al 1995, Henry et al 1995, Mulchaey et al
1996a). The narrow range of group masses is not too surprising, given
that nearly all the groups in these surveys have temperatures of $\sim$
1 keV. 

\begin{figure}
\psfig{figure=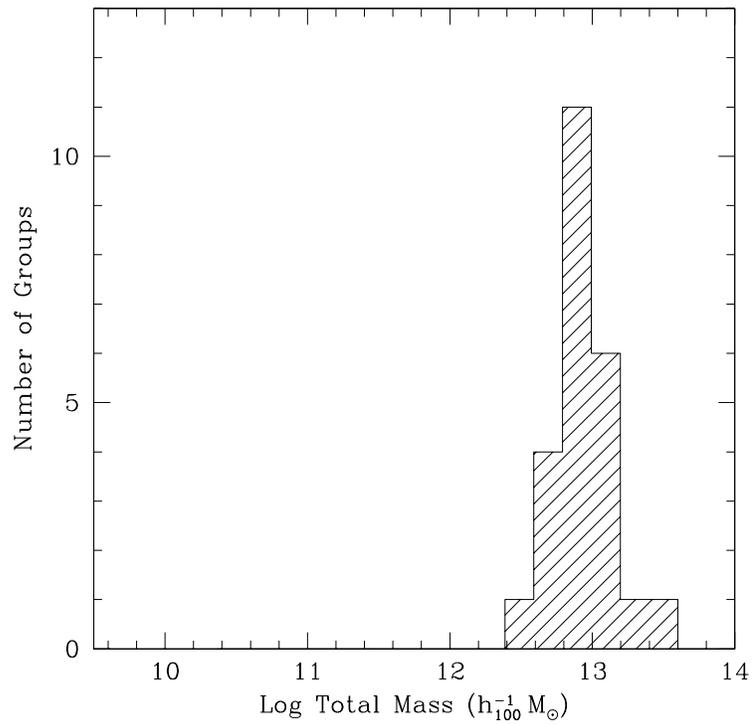,width=4.0in}
\caption{Distribution of X-ray--determined total group masses. 
In each case, the masses are determined out to the radius to which
the X-ray emission is detected.
The sample is based on the compilation given in Mulchaey et al 1996a,
with the addition of a few groups with more recent X-ray mass estimates 
in the literature.} 
\end{figure}

The X-ray mass estimates can generally be applied only to a radius of
several hundred kiloparsecs. Beyond that, the gas density profile is
not well-constrained. Because the virial radius for a 1 keV group is
approximately $\sim$ 0.5 h$_{\rm 100}$$^{-1}$ Mpc, the X-ray method
measures only a fraction of the total mass (Ponman \& Bertram 1993;
David et al 1995; Henry et al 1995). Simply extrapolating out to the virial
radius, the total group masses are a factor of approximately two to
three times larger than those implied from the X-ray studies (Mass
$\propto$ R). However, if non-gravitational heat is important in groups,
the extrapolation out to the virial radius is more uncertain (
Loewenstein 2000).

Because of their relatively large masses, X-ray groups make a substantial
contribution to the mass density of the universe (Mulchaey et al 1993,
Henry et al 1995, Mulchaey et al 1996a). Based on their X-ray selected
group sample, Henry et al (1995) estimate that X-ray luminous groups
contribute $\Omega$ $\sim$ 0.05. However, their sample contained only
the most luminous, elliptical-rich groups. When one corrects for the
groups missing from Henry et al's (1995) sample (assuming a similar
mass density), groups might contribute as much as $\Omega$ $\sim$
0.25. These estimates are comparable to the numbers found for richer
clusters, which verifies the cosmological significance of poor groups.

\subsection{Baryon Fraction}

The ratio of baryonic to total mass in groups and clusters can provide
interesting constraints on cosmological models (e.g. Walker et al
1991, White et al. 1993). The two known baryonic components in groups
are the galaxies and the hot gas. The total mass in galaxies can be
estimated by measuring the total galaxy light and assuming an
appropriate mass to light ratio for each galaxy based on its
morphological type. While ideally the luminosity function of each
group should be used to measure the total light, generally most authors have
included only the contribution of the most luminous
galaxies. Fortunately, these galaxies account for nearly all the light
in the group. The mass-to-light ratios of X-ray groups are generally
in the range M/L$_{\rm B}$ $\sim$ 120--200 h$_{\rm 100}$
M$_{\odot}$/L$_{\odot}$ (Mulchaey et al 1996a), which is comparable to
the mass-to-light ratios found in rich clusters. However, these
estimates are made out to the radius of X-ray detection, so the values
out to the virial radius could be larger. 
Assuming standard mass-to-light ratios for ellipticals and spirals,
the mass in galaxies in X-ray groups is typically in the range 
3 $\times$ 10$^{11}$--2 $\times$ 10$^{12}$
 h$_{\rm 100}$$^{-1}$ M$_{\odot}$
(Figure 9).

\begin{figure}
\psfig{figure=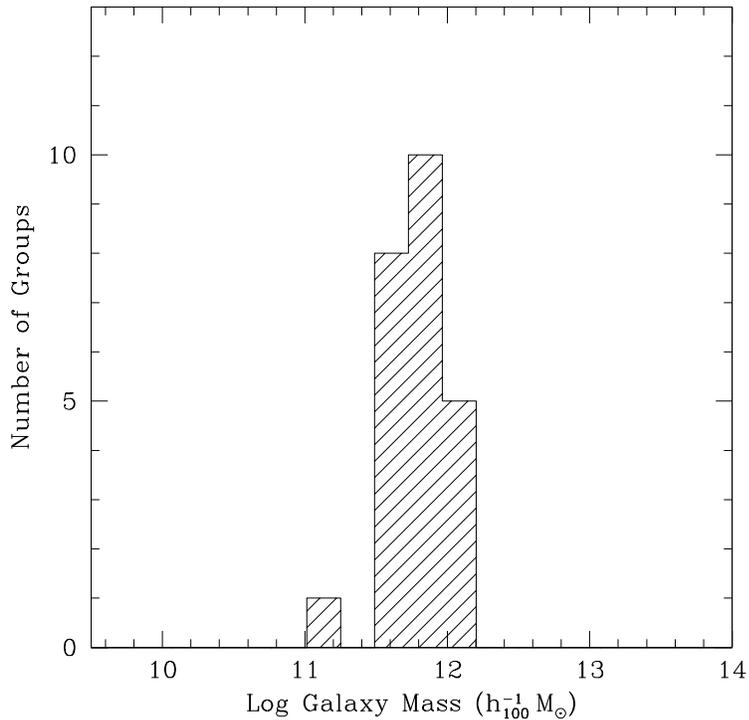,width=4.0in}
\caption{Distribution of galaxy mass for the sample of groups
used in Figure 8. 
}
\end{figure}

\begin{figure}
\psfig{figure=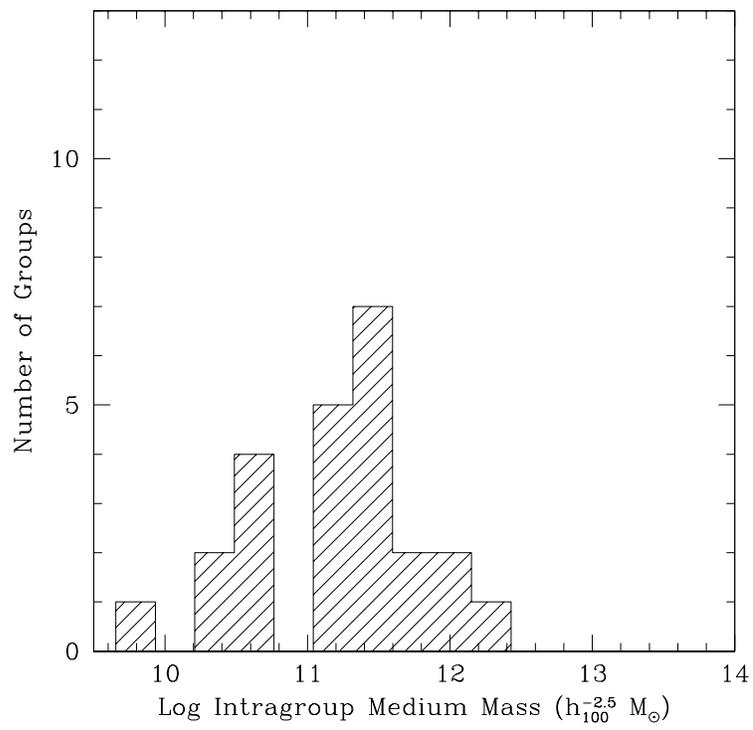,width=4.0in}
\caption{Distribution of intragroup medium mass for the sample of groups
used in Figure 8. 
}
\end{figure}

The mass in the intragroup medium can be estimated from the model fit
to the surface brightness profile. The gas mass estimates depend
on both the radius out to which X-rays are detected
(Henriksen \& Mamon 1994) and on the spectral properties assumed (for
example, the gas metallicity; Pildis et al 1995). For these reasons,
different authors have derived significantly different gas masses for
the same systems (cf Mulchaey et al 1996a). For most groups, the gas
mass is in the range $\sim$ 2 $\times$ 10$^{10}$--10$^{12}$ 
h$_{\rm 100}$$^{-5/2}$
M$_{\odot}$ (Figure 10). This is somewhat less than or
comparable to the mass in galaxies. Note, however, that the gas mass
is much more strongly dependent on H$_{\rm o}$, and for more realistic 
(i.e. lower) values of
H$_{\rm o}$, the gas mass can be somewhat higher than the galaxy
mass. The observed gas mass to stellar mass ratio tends to decrease as the
temperature of the system decreases. This trend extends from rich
clusters to individual elliptical galaxies. David et al (1995)
estimate that the gas to total mass fraction is approximately 2\% in
ellipticals, 10\% in groups and 20--30\% in rich clusters. 
However, the hot gas in groups is detected to a much smaller fraction of 
the virial radius than in
rich clusters, so comparisons made at the current level of X-ray detection
may not accurately reflect the global gas fractions (Loewenstein
2000). In fact, much of the intragroup gas probably lies beyond the current
X-ray 
detection limits, and on more global scales, groups may not be gas--poor
compared to clusters.
Consequently, the total gas masses of groups 
may be severely underestimated by ROSAT observations. On scales
of the virial radius, the intragroup
medium is likely the dominant baryonic component in these systems. 
In fact, Fukugita et al (1998) estimated
that diffuse gas in groups is the dominant baryon component in the
nearby universe. A fundamental assumption in Fukugita et al's
calculation is that all groups contain an intragroup medium and that
the absence of X-ray detections in many groups is primarily a result of
lower virial temperature rather than the absence of plasma. 
Regardless of whether this assumption is valid or not, it is now clear that
intragroup gas is an important baryonic
constituent of the local universe.

\begin{figure}
\psfig{figure=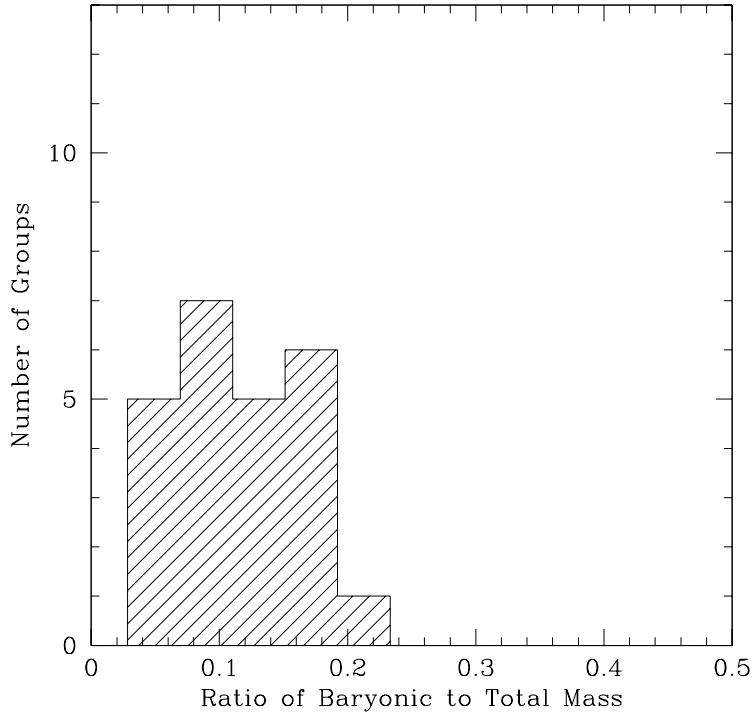,width=4.0in}
\caption{Distribution of total observed baryonic mass to total group mass
for the sample of groups
used in Figure 8. The low \lq\lq baryonic fractions\rq\rq \ derived for 
groups indicate 
that these systems are dominated by dark matter.
}
\end{figure}

Adding up the baryons in galaxies and intragroup gas and comparing to
the total mass, one finds that the known baryonic 
components typically account for only
10--20\% of the total mass that is derived using the X-ray data
(Figure 11;
 Mulchaey et al 1993; Ponman \& Bertram 1993; David et al 1994; Pildis
et al 1995; David et al 1995; Doe et al 1995; Davis et al 1995, 1996;
Mulchaey et al 1996a; Pedersen et al 1997). This provides some of the
strongest evidence to date that small groups of galaxies are dominated
by dark matter. The ratio of mass in observed baryonic components to
total mass (i.e. the \lq\lq baryon fraction\rq\rq ) in general is
smaller in groups than in rich clusters (David et al 1995, David
1997). 
However, the lower observed baryon fractions of groups may largely
reflect the fact that
much of the hot gas occurs beyond the
radius of current X-ray detection. 
Even if the observed baryon fractions of groups are representative of 
the global values,
the baryon fractions in X-ray groups are still too high to
be consistent with the low baryon fractions required for $\Omega$=1 and
standard big bang nucleosynthesis (cf White et al 1993).

\subsection{Large-Scale Structure}
Redshift surveys of the nearby universe indicate that groups of
galaxies are good tracers of large-scale structure (e.g. Ramella et al
1989). The presence of a hot intragroup medium in many groups
suggests that X-ray observations can also be used to map out the
distribution of mass in the universe. Recent ROSAT results demonstrate
the great potential of large area X-ray surveys. Mullis et al (2000)
have recently completed an optical follow-up survey of the $\sim$ 500
X-ray sources detected in the ROSAT All-Sky Survey in a 9 $\times$ 9
square-degree region around the north ecliptic pole. They identify 65
galaxy systems, $\sim$ 30\% of which are poor groups. Remarkably,
some 23\% of the galaxy systems found in this field belong to a single
wall-like structure at z = 0.088. Although a supercluster consisting of six
Abell clusters had previously been identified in this region (Batuski
\& Burns 1985), the X-ray data reveal that this supercluster is
significantly larger than implied by the optical data alone.
Furthermore, the X-ray data show that the massive Abell clusters are
linked together by groups and poor clusters. The supercluster spans
the entire area surveyed by Mullis et al (2000), suggesting that the true
extent of this structure could be larger still. Numerical simulations
imply that future X-ray missions such as CHANDRA and
 XMM will be able to map out
even lower-density regions such as filaments (Pierre et al 2000). Such
X-ray studies will be very important because many current models
suggest that the majority of baryons occur in these
filaments (Miralda-Escude et al 1996, Cen \& Ostriker 1999).

\subsection{Moderate Redshift Groups}

Despite the cosmological significance of groups, remarkably little is
known about these systems at high redshift. Optical studies of high
redshift groups have been limited because low galaxy densities make
groups difficult to recognize even at moderate redshifts. X-ray
emission from the intragroup medium provides a potentially useful
method for finding groups at high redshift. A number of searches for
faint, extended X-ray sources have been performed in recent years
using deep ROSAT PSPC observations (e.g. Rosati et al 1995, Griffiths
et al 1995, Scharf et al 1997, Burke et al 1997, Jones et al 1998,
Schmidt et al 1998, Vikhlinin et al 1998, Zamorani et al 1999).
Although the goal of these surveys is often to find rich clusters of
galaxies at high redshift, many X-ray groups at redshifts z=0.1--0.6
have also been found. Unfortunately, the ROSAT observations of these
groups generally contain very few counts, so it is not possible to
determine the temperature or the metallicity of the gas with the
existing data. However, studies of the spectral properties of the
intragroup medium out to z $\sim$ 0.3 will be possible with both XMM
and CHANDRA. Furthermore, deep images with these telescopes will
likely uncover X-ray groups at even higher redshifts. Therefore, the
first studies of the evolution of the intragroup medium should be
possible within the next decade.

\subsection{Gravitational Lensing}

The efficiency of a massive system to act as a gravitational lens is a
function of both the mass density profile and the source--lens--observer
geometry (cf Blandford \& Narayan 1992). Given their relatively high
mass densities, X-ray groups at moderate (z $>$ 0.2) redshifts are
expected to be efficient lenses (Mendes de Oliveira \& Giraud 1994,
Montoya et al 1996). Unfortunately, because very few samples of galaxy
groups at moderate redshift exist in the literature, systematic
searches for lensing in these objects have not been carried
out. However, several of the well-studied, multiple-image QSO systems
are lensed by galaxies that belong to spectroscopically-confirmed poor
groups (Kundic et al 1997a,b; Tonry 1998; Tonry \& Kochanek 2000).
Although the primary lens in each of these cases is an individual galaxy,
the group potential also contributes to the observed lensing. The
presence of an extended group potential acts as a source of external
sheer (Keeton et al 1997; Kundic et al 1997a,b). To properly model
the lensing system, the group potential must be included. Most
authors have attempted to measure the velocity dispersion of the group
and then assume a form for the potential. Unfortunately, these
dispersions are based on only a few velocity measurements and are
subject to the large uncertainties that have plagued optical studies
of nearby groups. Still, good fits to the lensing data are often
obtained. In the case of the quadruple lens PG 1115+080, the measured
velocity dispersion of the group (Kundic et al 1997a; Tonry 1998) is
consistent with the value predicted earlier from the lensing data
(Schecter et al 1997). Obtaining an adequate model for the group
potential is also necessary to derive cosmological parameters like the
Hubble Constant (H$_{\rm o}$) from lensing experiments. Future X-ray
observations may be the key to such techniques. High-resolution X-ray
images taken with CHANDRA and XMM should allow the potential of the
lensing groups to be mapped in detail. A better determination of the
lensing potential will result in tighter constraints on cosmological
parameters.

\subsection{Cooling Flows}
Galaxy groups display many of the signatures of cooling flows that
have previously been observed in rich clusters and elliptical galaxies
(Fabian 1994). The surface brightness profiles of the X-ray emission
are sharply peaked, indicating that the gas density is rising rapidly
towards the center of the group. In addition, at least half of all
groups with measured temperature profiles show direct evidence for
cooler gas in the central regions (Ponman \& Bertram 1993; David et al
1994; Trinchieri et al 1997; Mulchaey \& Zabludoff 1998; Helsdon \&
Ponman 2000). In some cases, the central gas is cooler than the mean
gas temperature by nearly 50\%. Cooling flow models also appear to
provide a better fit to the ASCA spectra of groups than an isothermal
plasma model (Buote 2000a). While these observations are consistent
with the cooling flow interpretation, there are other possibilities.
For example, Mulchaey \& Zabludoff (1998) noted that the above
features could also be explained if there is a distinct X-ray
component associated with the central elliptical galaxy.

Perhaps the strongest case for a cooling flow in a low-mass system is
the NGC 5044 group. David et al (1994) obtained a very deep ROSAT PSPC
observation of this system that allowed the construction of a detailed
temperature profile. They found evidence for a cooling flow with an
essentially constant mass accretion rate from approximately 20 h$_{\rm
100}$$^{-1}$ kpc out to the cooling radius ($\sim$ 50--75 h$_{\rm
100}$$^{-1}$ kpc). This suggests a nearly homogeneous cooling flow.
In contrast, the cooling flows in rich clusters tend to be
inhomogeneous; a significant amount of the gas cools out at
large radii (cf Fabian 1994). David et al (1994) suggest that
gravitational heating is more important in the NGC 5044 group than in
clusters because in groups the temperature of the hot gas is
comparable to the virial temperature of the central galaxy, whereas for
rich clusters the gas temperature is significantly higher. Therefore,
most of the observed X-ray emission in the cooling flow region can be
provided by the gravitational energy in groups, whereas mass deposition
dominates in rich clusters. 

\subsection{Fossil Groups}

Because of their relatively low velocity dispersions and high galaxy
densities, groups of galaxies provide ideal sites for galaxy-galaxy
mergers. Numerical simulations suggest that the luminous galaxies in a
group will eventually merge to form a single elliptical galaxy (Barnes
1989, Governato et al 1991, Bode et al 1993, Athanassoula et al
1997). The merging timescales for the brightest group members
(M $\simless$ M$^*$) are typically a few tenths of a Hubble time for
an X-ray detected group (Zabludoff \& Mulchaey 1998). Therefore,
by the present day some
groups have likely merged into giant ellipticals.
Outside of the high-density core, the cooling time for the intragroup
medium is longer than a Hubble time; thus, while the luminous
galaxies in some groups have had enough time to merge into a single
object, the large-scale X-ray halo of the original groups should
remain intact. This means that a merged group might appear today as an isolated
elliptical galaxy with a group-like X-ray halo (Ponman \& Bertram
1993).

Using the ROSAT All-Sky Survey data, Ponman et al (1994) found the
first such \lq\lq fossil\rq\rq\ group candidate. The RXJ1340.6+4018
system has an X-ray luminosity comparable to a group, but $\sim$ 70\%
of the optical light comes from a single elliptical galaxy (Jones et
al 2000). The galaxy luminosity function of RXJ1340.6+4018 indicates a
deficit of galaxies at approximately M$^*$. The luminosity of the
central galaxy is consistent with it being the merger product of the
missing M$^*$ galaxies. Jones et al (2000) have studied the central
galaxy in detail and find no evidence for spectral features implying
recent star formation, which indicates the last major merger occurred at
least several gigayears ago.

Several other fossil group candidates are now known. Mulchaey \&
Zabludoff (1999) discovered a large X-ray emitting halo around the
optically- selected isolated elliptical NGC 1132. Although the NGC 1132
system contains no other luminous galaxies, there is evidence for an
extensive dwarf galaxy population clustered around the central galaxy.
The dwarfs in NGC 1132 are comparable in number and distribution to
the dwarfs found in X-ray groups (Zabludoff \& Mulchaey 1998). The
existence of a clustered dwarf population in fossil groups is not
surprising because the galaxy-galaxy merger and dynamical friction
timescales for faint galaxies in groups are significantly longer than
the timescales for the luminous galaxies (Zabludoff \& Mulchaey 1998).
Hence, the dwarf galaxy population, like the X-ray halo, will remain
long after the central elliptical has formed.

Vikhlinin et al (1999)
have found four potential fossil groups in their large-area ROSAT
survey of extended X-ray sources. (Their sample includes RXJ1340.6+4018
and two X-ray sources detected in earlier Einstein surveys but not
previously recognized as potential group remnants.) Given the large
surface area they covered in their survey, Vikhlinin et al were able
to estimate the spatial density of X-ray fossil groups for the first time
and found that these objects represent $\sim$ 20\% of all clusters and
groups with an X-ray luminosity greater than 5 $\times$ 10$^{42}$
h$_{\rm 100}$$^{-2}$ ergs s$^{-1}$. The number density of fossil
groups is comparable to the number density of field ellipticals, so
most, if not all, luminous field ellipticals may be the product of
merged X-ray groups.

Although the X-ray and optical properties of some luminous, isolated
elliptical galaxies are consistent with the merged group
interpretation, another possibility is that these systems may have
simply formed with a deficit of luminous galaxies (Mulchaey \&
Zabludoff 1999). Distinguishing between these two scenarios will be
difficult, if not impossible. Regardless, these objects are massive
enough and found in large enough numbers that they are cosmologically
important. Vikhlinin et al (1999) estimated that the contribution of
fossil groups to the mass density of the universe is comparable to the
contribution of massive clusters. These objects are also an important
reminder that galaxies are not always a good tracer of mass and
large-scale structure: optical group catalogs would miss these large
mass concentrations.

\subsection{The Origin and Evolution of the Intragroup Medium} 

The presence of heavy elements in the intragroup medium indicates that
a substantial fraction of the diffuse gas must have passed through
stars. The presence of iron is particularly important because it
suggests that supernovae played an important role in the enrichment of
the gas. 
 In principle, X-ray spectroscopy can provide detailed
constraints on the stars responsible for the enrichment. For example,
the relative abundance of the $\alpha$-burning elements to iron is a
measure of the relative importance of Type II to Type 1a supernovae
(Renzini et al 1993, Renzini 1997, Gibson et al 1997). For the gas
temperatures characteristic of groups ($\sim$ 1 keV), strong emission
lines are expected for many of the $\alpha$ elements including oxygen,
neon, magnesium, silicon and sulfur. Although most ASCA studies
suggest that the $\alpha$/Fe ratio is approximately solar in groups, this
result is somewhat inconclusive at present because of
uncertainties in the spectral modeling. 

Renzini and collaborators have used the concept of iron mass-to-light
ratio to study the history of the hot gas in groups and clusters
(Renzini et al 1993, Renzini 1997). They find that the X-ray emitting
gas in rich clusters contains $\sim$ 0.01 h$^{-1/2}$ M$_{\odot}$ of
iron for each L$_{\odot}$ of blue light. The iron mass-to-light ratio
is effectively constant for clusters with temperature between $\sim$ 2
and 10 keV. However, this ratio is typically a factor of $\sim$ 50
lower in X-ray groups (Renzini et al 1993, Renzini 1997, Davis et al
1999). The iron mass-to-light ratios of groups are lower than those
of clusters because both the overall iron abundance and the gas to
stellar mass ratio are lower in groups than in clusters (Renzini 1997).
The low iron mass-to-light ratios may be evidence that a significant
amount of mass has been lost in groups. The escape velocities of
groups are comparable to the escape velocities of individual
galaxies. Thus, material that is ejected from galaxies may also escape
the group. Several mechanisms have been proposed to eject material
from groups, including galactic winds and outflows powered by
supernovae or nuclear activity (Renzini 1997). The material lost from
groups may have contributed significantly to the enrichment of the
intergalactic medium (Davis et al 1999).

The iron mass-to-light ratios of groups could be somewhat
underestimated if the true iron abundances are higher than the
sub-solar values usually derived from isothermal model fits. However,
the gas mass estimates are less sensitive to the iron abundance
assumed and uncertainties in the iron abundances likely lead to
inaccuracies in the gas mass estimates of at most $\sim$ 50\% (Pildis
et al 1995).
A potentially bigger problem is that many groups are
detected to a much smaller fraction of the virial radius than their
rich clusters counterparts. Thus, the true gas masses in some groups 
may be significantly underestimated from the existing X-ray 
data. In fact, it is possible that the differences in 
the iron mass-to-light ratios of groups and clusters may largely 
be a result of
this effect and not necessarily evidence for mass loss.

The mechanisms responsible for producing metals may also inject energy
into the gas. Numerical simulations indicate that in the absence of
such non-gravitational heating, the density profiles of groups and
clusters are nearly identical (Navarro et al 1997). There is now
considerable evidence for departures from such uniformity. In the
standard hierarchical clustering models, the X-ray luminosity is
expected to scale with temperature as L$_{\rm X}$ $\propto$ T$^{2}$
(e.g. Kaiser 1991). The observed relationship is considerably steeper,
especially for small groups (see Figure 6). Furthermore, the ratio of
specific energy of the galaxies to specific energy of the gas
(i.e. the $\beta$ parameter) is less than one for low-mass systems.
(However, see Section 3.3.3 for a discussion of why the observed $\beta$ values
for groups may be biased low).
Both of these observations suggest that the gas temperature may not
be a good indicator of the virial temperature in poor groups. Entropy
profiles for groups and clusters indicate that the entropy of the group
gas is also higher than can be achieved through gravitational collapse
alone (David et al 1996, Ponman et al 1999, Lloyd-Davies et al 2000).
All of these observations are consistent with preheating models for
the hot gas (Kaiser 1991; Evrard \& Henry 1991; Metzler \& Evrard
1994; Knight \& Ponman 1997; Cavaliere et al 1997, 1998, 1999; Arnaud
\& Evrard 1999; Balogh et al 1999; 
Tozzi et al 2000; Loewenstein 2000;
Tozzi \& Norman 2000).
Such preheating leads to a more extended gas component in groups than
in rich clusters (i.e. lower central gas densities and shallower 
density slopes).
Moreover, without preheating, groups appear to over-produce the X-ray
background (Wu et al 2000).

Ponman and collaborators have estimated the excess entropy associated
with the preheating in groups and find that it corresponds to a
temperature of $\sim$ 0.3 keV (Ponman et al 1999, Lloyd-Davies et al
2000). The preheating temperature can be combined with the excess
entropy to estimate the electron density of the gas into which the
energy was injected. The resulting value (n $\sim$ 4 $\times$
10$^{-4}$ h$_{\rm 100}$$^{0.5}$ cm$^{-3}$) implies that the heating
occurred prior to the cluster collapse but after a redshift of z
$\sim$ 10 (Lloyd-Davies et al 2000). The current estimates for the
entropy associated with the preheating have been based on rather small
samples of groups and clusters, and these techniques will undoubtably
improve with the next generation of X-ray telescopes. Already it is
clear that such research can provide considerable insight into the
history of the gas and group formation.

\subsection{The Local Group}
Finally, it is interesting to consider the implications X-ray 
observations of other groups have for our own Local Group.
The idea that the Local Group might contain a hot intragroup medium
dates back to the work of Kahn \& Woltjer (1959). The X-ray detection
of other groups has led to renewed interest in 
this idea. Suto et al (1996) proposed
that a hot halo around the Local Group with a temperature of $\sim$ 1
keV and column density N$_{\rm H}$ $\sim$ 10$^{21}$ cm$^{-2}$ could
explain the observed excess in the X-ray background below 2 keV. The
X-ray halo would also generate temperature anisotropies in the
microwave background via the Sunyaev-Zeldovich effect. There is no
evidence for such anisotropies in the COBE MDR maps, however (Banday
\& G\'{o}rski 1996). Furthermore, the gas temperature and column
density assumed by Suto et al (1996) are probably overestimated given
the ROSAT observations of other groups (Pildis \& McGaugh 1996). In
fact, the strong trend for spiral-only groups not to be X-ray detected
suggests that the Local Group is unlikely to produce appreciable amounts of
X-ray emission (Pildis \& McGaugh 1996, Mulchaey et al 1996b).

Although the Local Group is probably not X-ray bright, a significant
gas component may exist at cooler temperatures (Mulchaey et al 1996b,
Fields et al 1997). Given the expected virial temperature of the Local
Group ($\sim$ 0.2 keV), the detection of this gas in emission would be
exceedingly difficult. However, an enriched collisionally ionized gas
at these temperatures is expected to produce prominent absorption
features in the far-UV region. The strongest features result from
lithium-like ions O VI, Ne VIII, Mg X and Si XII (Verner et al 1994).
Lines of sight to hot stars in the Magellanic Clouds are known to show
O VI absorption features, but it is not clear whether this gas is
associated with intragroup gas or gas in our own Galaxy. There may be
other ways to infer the presence of warm gas in the Local Group. Wang
\& McCray (1993) found evidence in the soft X-ray background for a
thermal component with temperature $\sim$ 0.2 keV, which could be due
to a warm intragroup medium in the Local Group (see, however, Sidher
et al 1999, who argue that the X-ray halo of the Galaxy dominates).
Maloney \& Bland-Hawthorn (1999) have recently considered the ionizing
flux produced by warm intragroup gas and find that it is unlikely to
dominate over the cosmic background or the ultraviolet background
produced by the luminous members of the Local Group. Still, encounters
between the intragroup gas and the Magellanic Stream may be
responsible for the strong H$\alpha$ emission detected by Weiner \&
Williams (1996).

The existence of an intragroup medium in the Local Group may also be 
relevant to the H I high velocity clouds (HVCs; for a review see 
Wakker \& van Woerden 1997). Recently, Blitz et al (1999) revived
the idea that many of the HVCs may be dark-matter dominated structures
falling onto the Local Group. In this scenario, some of the HVCs 
collide near the center of the Local Group and produce a warm 
intragroup medium. If the Blitz et al (1999) scenario is 
correct, one would expect to
find similar H I clouds in other nearby groups. Blitz et al (1999)
suggested that several HVC analogs have indeed been found. 
However, Zwaan \& Briggs (2000) completed a
H I strip survey of the extragalactic sky with Arecibo and detected no
objects resembling the HVCs in other groups. 
The failure of the Arecibo survey to 
detect H I does not necessarily rule out the Blitz et al (1999) model.
One possibility is that
the groups in the Zwaan \& Briggs (2000) survey contain an 
X-ray emitting intragroup gas and that the H I clouds do not survive 
this hostile environment. Unfortunately, the X-ray properties of the
Zwaan \& Briggs (2000) groups are currently unknown. 
The conclusions of Zwaan \& Briggs (2000) are also sensitive to the 
masses assumed for the H I clouds. 
Braun \& Burton (2000) argued for a lower HVC H I mass and concluded that 
the sensitivity and coverage of Zwaan \& Briggs' (2000)
survey was not sufficient to detect analogs of the HVCs in other groups.
A more serious problem may be the number statistics of moderate redshift
Mg II and Lyman limit absorbers, which appear to be inconsistent with
a Local Group origin for the HVCs (Charlton et al 2000). Regardless, 
it is clear that future H I surveys of X-ray detected and X-ray--non-detected 
groups could provide important insight into the relationship between 
hot and cold gas in galaxy groups.

\section{FUTURE WORK}
X-ray telescopes launched in the 1990s have firmly established the
presence of a hot X-ray emitting intragroup medium in nearby groups of
galaxies. X-ray observations suggest that many groups are real,
physical systems. The masses of X-ray groups are substantial and make
a significant contribution to the mass density of the
universe. Although most of the mass in groups appears to be in dark
matter, the intragroup medium may be the dominant baryonic component
in the nearby universe.

While we have made significant progress towards understanding groups
in the last decade, there are still many outstanding issues. Ambiguities
about the proper spectral model for the gas and 
our inability to detect gas to a large fraction of the virial radius
 are particularly troubling
because the resulting uncertainties propagate into cosmological
applications. Furthermore, the contribution of individual galaxies to
the observed X-ray emission remains a point of contention. Our ability
to understand the intragroup medium has largely been limited by the poor
spatial and spectral resolution of the X-ray instruments. This
situation is about to change drastically, however, with the
availability of new powerful X-ray telescopes. Recently, NASA
successfully launched CHANDRA (formerly known as AXAF). This telescope
will produce high-resolution X-ray images of groups ($\sim$ 1$''$)
that will allow the relative contribution of galaxies and diffuse gas
to be quantified. In late 1999, the European Space Agency (ESA) 
launched XMM-Newton. Although 
the spatial resolution of XMM-Newton is poorer than that of
CHANDRA, the collecting area of this telescope is much
greater. Therefore, XMM-Newton will obtain the deepest X-ray exposures ever
of nearby groups and will extend the studies of the group environment to
higher redshifts. 
The combination of CHANDRA and XMM-Newton will probably answer
many of the questions raised by the recent generation of X-ray
telescopes.

\vskip 0.2cm
I would like to thank my collaborators and colleagues 
particularly  Arif Babul, Dave Burstein, 
David Buote, David Davis,
Steve Helsdon, Pat Henry, Lawrence Jones, Lori Lubin, Gary Mamon,
Bill Mathews, Kyoko Matsushita,
Chris Mullis, Richard Mushotzky, Gus Oemler, Trevor Ponman, 
Matthias Steinmetz,
Jack
Sulentic, Ben Weiner, and Ann Zabludoff for useful
discussions on X-ray groups. I would also like to thank
David Davis, Richard Mushotzky and Allan Sandage for comments on 
the manuscript.
 This work was supported in part by NASA
under grant NAG 5-3529.

\vfill\eject

\listoffigures


\begin{thebibliography}{99}

\bibitem{}Albert CE, White RA, Morgan WW. 1977.  {\it Ap. J.}
211:309--10

\bibitem{}Allen SW, Fabian AC, Edge AC, B\"{o}hringer H, White DA. 1995.
{\it MNRAS} 275:741--54

\bibitem{}Arimoto N, Matsushita K, Ishimaru Y, Ohashi T, Renzini A. 1997
{\it Ap. J.} 477:128--143

\bibitem{}Arnaud M, Evrard AE. 1999.
{\it MNRAS} 305:631--40

\bibitem{}Aschenbach B. 1988.
{\it Appl. Optics} 27:1404--13

\bibitem{}Athanassoula E, Makino J, Bosma A. 1997.
{\it MNRAS} 286:825--38 

\bibitem{}Bahcall NA, Harris DE, Rood HJ. 1984.
{\it Ap. J.} 284:L29--33

\bibitem{}Bahcall NA, Lubin LM. 1994.
{\it Ap. J.} 426:513--15

\bibitem{}Balogh ML, Babul A, Patton DR. 1999.
{\it MNRAS} 307:463--79

\bibitem{}Banday AJ, G\'{o}rski KM. 1996.
{\it MNRAS} 283:L21--25

\bibitem{}Barnes JE. 1985.
{\it MNRAS} 215:517--36

\bibitem{}Barnes JE. 1989.
{\it Nature} 338:123--26

\bibitem{}Bartelmann M, Steinmetz M. 1996.
{\it MNRAS} 283:431--46

\bibitem{}Batuski DJ, Burns JO. 1985.
{\it Astron. J.} 90:1413--24

\bibitem{}Bauer F, Bregman JN. 1996.
{\it Ap. J.} 457:382--89

\bibitem{}Bergeron J, Petitjean P, Sargent WLW, Bahcall JN, Boksenberg A, 
et al. 1994.
{\it Ap. J.} 436:33--43

\bibitem{}Biermann P, Kronberg PP. 1983.
{\it Ap. J.} 268:L69--73

\bibitem{}Biermann P, Kronberg PP, Madore BF. 1982.
{\it Ap. J.} 256:L37--40

\bibitem{}Bird CM, Mushotzky RF, Metzler CA. 1995.
{\it Ap. J.} 453:40--47

\bibitem{}Blandford RD, Narayan R. 1992.
{\it Annu. Rev. Astron. Astrophy.} 30:311--58

\bibitem{}Blitz L, Spergel DN, Teuben PJ, Hartmann D, Burton WB. 1999.
{\it Ap. J.} 514:818--43

\bibitem{}Bode PW, Cohn HN, Lugger PM. 1993.
{\it Ap. J.} 416:17--25

\bibitem{}Braun R, Burton WB. 2000.
{\it Astron. Astrophys.} submitted

\bibitem{}Buote DA. 1999.
{\it MNRAS} 309:685--714

\bibitem{}Buote DA. 2000a.
{\it MNRAS} 311:176--200

\bibitem{}Buote DA. 2000b.
{\it Ap. J.} 532:L113--116

\bibitem{}Buote DA. 2000c.
{\it Ap. J.} 539:172--186

\bibitem{}Burke DJ, Collins CA, Sharples RM, Romer AK, Holden BP, et al. 1997.
{\it Ap. J.} 488:L83--86

\bibitem{}Burns JO, Gregory SA, Holman GD. 1981.
{\it Ap. J.} 250:450--63

\bibitem{}Burns JO, Ledlow MJ, Loken C, Klypin A, Voges W, et al. 1996. 
{\it Ap. J.} 467:L49--52

\bibitem{}Cavaliere A, Menci N, Tozzi P. 1997.
{\it Ap. J.} 484:L21--24

\bibitem{}Cavaliere A, Menci N, Tozzi P. 1998.
{\it Ap. J.} 501:493--508

\bibitem{}Cavaliere A, Menci N, Tozzi P. 1999.
{\it MNRAS} 308:599--608

\bibitem{}Cen R, Ostriker JP. 1999.
{\it Ap. J.} 514:1--6

\bibitem{}Charlton JC, Churchill CW, Rigby JR. 2000.
{\it Ap. J.} submitted

\bibitem{}Cooke BA, Ricketts MJ, Maccacaro T, Pye JP, Elvis M, et al. 1978.
{\it MNRAS} 182:489--515

\bibitem{}Cowie LL, Henriksen M, Mushotzky R. 1987.
{\it Ap. J.} 317:593--600

\bibitem{}David LP. 1997.
{\it Ap. J.} 484:L11--15

\bibitem{}David LP, Arnaud KA, Forman W, Jones C. 1990.
{\it Ap. J.} 356:32--40

\bibitem{}David LP, Jones C, Forman W. 1995.
{\it Ap. J.} 445:578--90

\bibitem{}David LP, Jones C, Forman W. 1996.
{\it Ap. J.} 473:692--706

\bibitem{}David LP, Jones C, Forman W, Daines S. 1994.
{\it Ap. J.} 428:544--54

\bibitem{}Davis DS, Mulchaey JS, Mushotzky RF. 1999. 
{\it Ap. J.} 511:34--40

\bibitem{}Davis DS, Mulchaey JS, Mushotzky RF, Burstein D. 1996.
{\it Ap. J.} 460:601--11

\bibitem{}Davis DS, Mushotzky RF, Mulchaey JS, Worrall DM, Birkinshaw M,
Burstein D. 1995.
{\it Ap. J.} 444:582--89

\bibitem{}Dell'Antonio IP, Geller MJ, Fabricant DG. 1994.
{\it Astron. J.} 107:427--47

\bibitem{}de Vaucouleurs G. 1965.
in {\it Stars and Setllar Systems}, ed. A. Sandage, M. Sandage and
J. Kristian (Chicago: Univesity of Chicago Press) 

\bibitem{}Diaferio A, Geller MJ, Ramella M. 1994.
{\it Astron. J.} 107:868--79

\bibitem{}Diaferio A, Geller MJ, Ramella M. 1995.
{\it Astron. J.} 109:2293--2303

\bibitem{}Diaferio A, Ramella M, Geller MJ, Ferrari A. 1993.
{\it Astron. J.} 105:2035--46

\bibitem{}Doe SM, Ledlow MJ, Burns JO, White RA. 1995.
{\it Astron. J.} 110:46--67

\bibitem{}Dos Santos S, Mamon GA. 1999.
{\it Astron. Astrophys.} 352:1--18 

\bibitem{}Ebeling H, Voges W, B\"{o}hringer H. 1994.
{\it Ap. J.} 436:44--55

\bibitem{}Edge AC, Stewart GC. 1991.
{\it MNRAS} 252:428--41  

\bibitem{}Evrard AE, Henry JP. 1991.
{\it Ap. J.} 383:95--103

\bibitem{}Evrard AE, Metzler CA, Navarro JF. 1996.
{\it Ap. J.} 469:494--507

\bibitem{}Fabian AC. 1994.
{\it Annu. Rev. Astron. Astrophys.} 32:277--318 

\bibitem{}Fabian AC, Arnaud KA, Bautz MW, Tawara Y. 1994.
{\it Ap. J.} 436:L63--66

\bibitem{}Fabricant D, Gorenstein P. 1983.
{\it Ap. J.} 267:535--46

\bibitem{}Fabricant D, Lecar M, Gorenstein P. 1980.
{\it Ap. J.} 241:552--60 

\bibitem{}Fabricant D, Rybicki G, Gorenstein P. 1984.
{\it Ap. J.} 286:186--95

\bibitem{}Fields BD, Mathews GJ, Schramm DN. 1997.
{\it Ap. J.} 483:625--37

\bibitem{}Finoguenov A, Ponman TJ. 1999.
{\it MNRAS} 305:325--37 

\bibitem{}Forman W, Jones C. 1982.
{\it Annu. Rev. Astron. Astrophys.} 20:547--85

\bibitem{}Frederic JJ. 1995.
{\it Ap. J. Suppl.} 97:259--74

\bibitem{}Fukazawa Y, Makishima K,
 Matsushita K, Yamasaki N, Ohashi T, et al. 1996.
{\it Publ. Astron. Soc. Japan} 48:395--407

\bibitem{}Fukazawa Y, Makishima K, Tamura T, Ezawa H, Xu H, et al. 1998.
{\it Publ. Astron. Soc. Japan} 50:187--93

\bibitem{}Fukugita M, Hogan CJ, Peebles PJE. 1998.
{\it Ap. J.} 503:518--30

\bibitem{}Geller MJ, Huchra JP. 1983.
{\it Ap. J. Suppl.} 52:61--87

\bibitem{}Gendreau KC. 1995.
{\it X-ray CCDs for Space Applications: Calibration, Radiation Hardness, 
and Use for Measuring the Spectrum of the Cosmic X-ray Background.} 
PhD thesis. Massachusetts Institute of Technology. 1 pp.

\bibitem{}Gibson BK, Loewenstein M, Mushotzky RF. 1997.
{\it MNRAS} 290:623--28

\bibitem{}Governato F, Bhatia R, Chincarini G. 1991.
{\it Ap. J.} 371:L15--18

\bibitem{}Governato F, Tozzi P, Cavaliere A. 1996.
{\it Ap. J.} 458:18--26

\bibitem{}Griffiths RE, Georgantopoulos I, Boyle BJ, Stewart GC, Shanks T, 
Della Ceca R. 1995.
{\it MNRAS} 275:77--88

\bibitem{}Griffiths RE, Schwartz DA, Schwarz J, Doxsey RE, Johnson MD, et al.
1979.
{\it Ap. J.} 230: L21--25

\bibitem{}Hellsten U, Gnedin NY, Miralda-Escude J. 1998.
{\it Ap. J.} 509:56--61

\bibitem{}Helsdon SF, Ponman TJ. 2000.
{\it MNRAS} 315:356--370

\bibitem{}Henriksen MJ, Mamon GA. 1994. 
{\it Ap. J.} 421:L63--66

\bibitem{}Henry JP, Gioia IM, Huchra JP, Burg R, McLean B, et al. 1995.
{\it Ap. J.} 449:422--30

\bibitem{}Hernquist L, Katz N, Weinberg DH. 1995.
{\it Ap. J.} 442:57--60

\bibitem{}Hickson P. 1982. 
{\it Ap. J.} 255:382--91

\bibitem{}Hickson P. 1997.
{\it Annu. Rev. Astron. Astrophys.} 35:357--88

\bibitem{}Hickson P, Huchra J, Kindl E. 1988.
{\it Ap. J.} 331:64--70

\bibitem{}Holmberg E. 1950.
{\it Medd. Lunds Obs. Ser. 2} 128:1--56

\bibitem{}Horner DJ, Mushotzky RF, Scharf CA. 1999.
{\it Ap. J.} 520:78--86

\bibitem{}Huchra JP, Geller MJ. 1982.
{\it Ap. J.} 257:423--37

\bibitem{}Humason ML, Mayall NU, Sandage AR. 1956.
{\it Ap. J.} 61:97--162

\bibitem{}Hunt R, Sciama DW. 1972.
{\it MNRAS} 157:335--48

\bibitem{}Hwang U, Mushotzky RF, Burns JO, Fukazawa Y, White RA. 1999.
{\it Ap. J.} 516:604--18

\bibitem{}Hwang U, Mushotzky RF, Loewenstein M, Markert, TH, Fukazawa Y,
 Matsumoto H. 1997.
{\it Ap. J.} 476:560--71

\bibitem{}Ikebe Y, Ezawa H, Fukazawa Y,
Hirayama M, Izhisaki Y, et al. 1996.
{\it Nature} 379:427--29

\bibitem{}Ishimaru Y, Arimoto N. 1997.
{\it PASJ} 49:1--8

\bibitem{}Jones C, Forman W. 1984.
{\it Ap. J.} 276:38--55

\bibitem{}Jones LR, Ponman TJ, Forbes DA. 2000.
{\it MNRAS} 312:139--50

\bibitem{}Jones LR, Scharf C, Ebeling H, Perlman E, Wegner G, et al. 1998.
{\it Ap. J.} 495:100--14 

\bibitem{}Kaastra JS, Mewe R. 1993.
{\it Astron. Astrophys. Suppl.} 97:443--82

\bibitem{}Kahn FD, Woltjer L. 1959.
{\it Ap. J.} 130:705--17

\bibitem{}Kaiser N. 1991.
{\it Ap. J.} 383:104--111

\bibitem{}Keeton CR, Kochanek CS, Seljak U. 1997.
{\it Ap. J.} 482:604--20

\bibitem{}King IR. 1962.
{\it Astron. J.} 67:471--85

\bibitem{}Knight PA, Ponman TJ. 1997.
{\it MNRAS} 289:955--72

\bibitem{}Kriss GA, Canizares CR, McClintock JE, Feidelson ED. 1980.
{\it Ap. J.} 235:L61--65

\bibitem{}Kriss GA, Cioffi DF, Canizares CR. 1983.
{\it Ap. J.} 272:439--48

\bibitem{}Kundic T, Cohen JG, Blandford RD, Lubin LM. 1997a.
{\it Astron. J.} 114:507--10

\bibitem{}Kundic T, Hogg DW, Blandford RD, Cohen JG, Lubin LM, et al. 1997b.
{\it Astron. J.} 114:2276--83

\bibitem{}Ledlow MJ, Loken C, Burns JO, Hill JM, White RA. 1996.
{\it Astron. J.} 112:388--406

\bibitem{}Liedahl DA, Kahn SM, Osterheld AL, Goldstein WH. 1990.
{\it Ap. J.} 350:L37--40

\bibitem{}Liedahl DA, Osterheld AL, Goldstein WH. 1995.
{\it Ap. J.} 438:L115--118

\bibitem{}Lloyd-Davies EJ, Ponman TJ, Cannon DB. 2000.
{\it MNRAS} In press

\bibitem{}Loewenstein M. 2000.
{\it Ap. J.} In press

\bibitem{}Mahdavi A, B\"{o}hringer H, Geller MJ, Ramella M. 1997.
{\it Ap. J.} 483:68--76

\bibitem{}Mahdavi A, Geller MJ, B\"{o}hringer H, Kurtz MJ, Ramella M. 1999.
{\it Ap. J.} 518:69--93

\bibitem{}Mahdavi A, B\"{o}hringer H, Geller MJ, Ramella M. 2000.
{\it Ap. J.} 534:114--132

\bibitem{}Maloney PR, Bland-Hawthorn J. 1999.
{\it Ap. J.} 522:L81--84

\bibitem{}Mamon GA. 1986.
{\it Ap. J.} 307:426--30

\bibitem{}Materne J. 1979.
{\it Astron. Astrophys.} 74:235--43

\bibitem{}Matsushita K, Ohashi T, Makishima K. 2000.
{\it PASJ} in press 

\bibitem{}McWilliam A. 1997.
{\it Annu. Rev. Astron. Astrophys.} 35:503--56

\bibitem{}Mendes de Oliveira C, Giraud E. 1994.
{\it Ap. J.} 437:L103--106

\bibitem{}Metzler CA, Evrard AE. 1994.
{\it Ap. J.} 437:564--83

\bibitem{}Mewe R. 1991.
{\it Astron. \& Astrophys. Review} 3:127--68

\bibitem{}Mewe R, Gronenschild EHBM, van den Oord GHJ. 1985.
{\it Astron. Astrophys. Suppl.} 62:197--254

\bibitem{}Miralda-Escude J, Cen R, Ostriker JP, Rauch M. 1996.
{\it Ap. J.} 471:582--616

\bibitem{}Mohr JJ, Evrard AE. 1997.
{\it Ap. J.} 491:38--44

\bibitem{}Mohr JJ, Mathiesen B, Evrard AE. 1999.
{\it Ap. J.} 517:627--49

\bibitem{}Montoya ML, Dominguez-Tenreiro R, Gonzalez-Casado G,
Mamon GA, Salvador-Sole E. 1996.
{\it Ap. J.} 473:L83--86

\bibitem{}Morgan WW, Kayser S, White RA. 1975.
{\it Ap. J.} 199:545--48 

\bibitem{}Morrison R, McCammon D. 1983.
{\it Ap. J.} 270:119--22

\bibitem{}Mulchaey JS, Davis DS, Mushotzky RF, Burstein D. 1993.
{\it Ap. J.} 404:L9--12

\bibitem{}Mulchaey JS, Davis DS, Mushotzky RF, Burstein D. 1996a.
{\it Ap. J.} 456:80--97

\bibitem{}Mulchaey JS, Davis DS, Mushotzky RF, Burstein D. 2000.
{\it Ap. J.} in preparation

\bibitem{}Mulchaey JS, Mushotzky RF, Burstein D, Davis DS. 1996b.
{\it Ap. J.} 456:L5--8

\bibitem{}Mulchaey JS, Zabludoff AI. 1998.
{\it Ap. J.} 496:73--92

\bibitem{}Mulchaey JS, Zabludoff AI. 1999.
{\it Ap. J.} 514:133--37

\bibitem{}Mullis CR, Henry JP, Gioia IM, B\"{o}hringer H, Briel UG. 2000.
{\it Ap. J.} in preparation.

\bibitem{}Mushotzky RF. 1984.
{\it Physica Scripta} T7:157--62

\bibitem{}Navarro JF, Frenk CS, White SDM. 1995.
{\it MNRAS} 275:720--40

\bibitem{}Navarro JF, Frenk CS, White SDM. 1997.
{\it Ap. J.} 490:493--508

\bibitem{}Nolthenius R. 1993.
{\it Ap. J. Suppl.} 85:1--25

\bibitem{}Nolthenius R, White SDM. 1987.
{\it MNRAS} 225:505--30

\bibitem{}Ohashi T, Ebisawa K, Fukazawa Y, Hiyoshi K, Horii M, et al. 1996.
{\it PASJ} 48:157--70

\bibitem{}Oort JH. 1970.
{\it Astron. Astrophys.} 7:381--404

\bibitem{}Ostriker JP, Lubin LM, Hernquist L. 1995.
{\it Ap. J.} 444:L61--64

\bibitem{}Pedersen K, Yoshii Y, Sommer-Larsen J. 1997.
{\it Ap. J.} 485:L17--20

\bibitem{}Perna R, Loeb A. 1998.
{\it Ap. J.} 503:L135--138

\bibitem{}Pfeffermann E, Briel UG, Hippmann H, Kettenring G, Metzner G, et al.
1988.
{\it Proc. SPIE} 733:519--32

\bibitem{}Pierre M, Bryan G, Gastaud R. 2000.
{\it Astron. Astrophys.} Submitted

\bibitem{}Pietsch W, Trinchieri G, Arp H, Sulentic JW. 1997.
{\it Astron. Astrophys.} 322:89--97

\bibitem{}Pildis RA, Bregman JN, Evrard AE. 1995.
{\it Ap. J.} 443:514--26

\bibitem{}Pildis RA, Evrard AE, Bregman JN. 1996.
{\it Astron. J.} 112:378--87

\bibitem{}Pildis RA, McGaugh SS. 1996.
{\it Ap. J.} 470:L77--79

\bibitem{}Ponman TJ, Allan DJ, Jones LR, Merrifield M, McHardy IM, 
et al. 1994.
{\it Nature} 369:462--64

\bibitem{}Ponman TJ, Bertram D. 1993.
{\it Nature} 363:51--54 

\bibitem{}Ponman TJ, Bourner PDJ, Ebeling H, B\"{o}hringer H. 1996.
{\it MNRAS} 283:690--708

\bibitem{}Ponman TJ, Cannon DB, Navarro JF. 1999.
{\it Nature} 397:135--137

\bibitem{}Price R, Duric N, Burns JO, Newberry MV. 1991.
{\it Astron. J.} 102:14--29

\bibitem{}Ramella M, Geller MJ, Huchra JP. 1989.
{\it Ap. J.} 344:57--74

\bibitem{}Ramella M, Geller MJ, Huchra JP, Thorstensen JR. 1995.
{\it Astron. J.} 109:1458--75

\bibitem{}Ramella M, Pisani A, Geller MJ. 1997.
{\it Astron. J.} 113:483--91

\bibitem{}Raymond JC, Smith BW. 1977.
{\it Ap. J. Suppl.} 35:419--39

\bibitem{}Renzini A. 1997.
{\it Ap. J.} 488:35--43

\bibitem{}Renzini A, Ciotti L, D'Ercole A, Pellegrini S. 1993.
{\it Ap. J.} 419:52--65

\bibitem{}Rhee G, van Haarlem M, Katgert P. 1992.
{\it Astron. J.} 103, 1721--28

\bibitem{}Ribeiro ALB, De Carvalho RR, Coziol R,
Capelato HV, Zepf SE. 1996.
{\it Ap. J.} 463:L5--8

\bibitem{}Ricker G, Doxsey RE, Dower RG, Jernigan JG, Delvailee JP, et al. 
1978. {\it Nature} 271:35--37

\bibitem{}Rosati P, Della Ceca R, Burg R, Norman C, Giacconi R. 1995.
{\it Ap. J.} 445:L11--14

\bibitem{}Ruderman MA, Spiegel EA. 1971.
{\it Ap. J.} 165: 1--15

\bibitem{}Saracco P, Ciliegi P. 1995.
{\it Astron. Astrophys.} 301:348--58

\bibitem{}Sarazin CL. 1986.
{\it Review Modern Physics} 58:1--115

\bibitem{}Sarazin CL, Burns JO, Roettiger K, McNamara BR. 1995.
{\it Ap. J.} 447:559--71

\bibitem{}Savage BD, Tripp TM, Lu L. 1998.
{\it Astron. J.} 115:436--50

\bibitem{}Scharf CA, Jones LR, Ebeling H, Perlman E, Malkan M, et al. 1997.
{\it Ap. J.} 477:79--92 

\bibitem{}Schechter PL, Bailyn CD, Barr R, Barvainis R, Becker CM,
et al. 1997.
{\it Ap. J.} 475:L85--88

\bibitem{}Schmidt M, Hasinger G, Gunn, J, Schneider D, Burg R, et al. 1998.
{\it Astron. Astrophys.} 329:495--503

\bibitem{}Schwartz DA, Schwarz J, Tucker W. 1980.
{\it Ap. J.} 238:L59--62

\bibitem{}Sidher SD, Sumner TJ, Quenby JJ. 1999.
{\it Astron. Astrophys.} 344:333--41

\bibitem{}Silk J, Tarter J. 1973.
{\it Ap. J.} 183:387--410

\bibitem{}Sulentic JW, Pietsch, Arp H. 1995.
{\it Astron. Astrophys.} 298:420--26

\bibitem{}Suto Y, Makishima K, Ishisaki Y, Ogasaka Y. 1996.
{\it Ap. J.} 461:L33--36

\bibitem{}Tanaka Y, Inoue H, Holt SS. 1994.
{\it Publ. Astron. Soc. Japan} 46, L37--41

\bibitem{}Tonry JL. 1998.
{\it Astron. J.} 115:1--5

\bibitem{}Tonry JL, Kochanek CS. 2000.
{\it Astron. J.} 117:2034--38

\bibitem{}Tozzi P, Norman C. 2000.
{\it Ap. J.} in press

\bibitem{}Tozzi P, Scharf C, Norman C. 2000.
{\it Ap. J.} in press

\bibitem{}Trinchieri G, Fabbiano G, Kim D-W. 1997.
{\it Astron. Astrophys.} 318:361--75

\bibitem{}Tully RB. 1987.
{\it Ap. J.} 321:280--304

\bibitem{}Verner DA, Tytler D, Barthel PD. 1994.
{\it Ap. J.} 430:186--90

\bibitem{} Vikhlinin A, McNamara BR, Forman W, Jones C, Quintana H, et al. 1998.
{\it Ap. J.} 502:558--81

\bibitem{} Vikhlinin A,
 McNamara BR, Hornstrup A, Quintana H, Forman W, et al. 1999.
{\it Ap. J.} 520:L1--4

\bibitem{}Voges W. 1993.
{\it Adv. Space Res.} 13:12391--97

\bibitem{}Wakker BP, van Woerden H. 1997.
{\it Annu. Rev. Astron. Astrophys.} 35:217--66

\bibitem{}Walke DG, Mamon GA. 1989.
{\it Astron. Astrophys.} 225:291--302

\bibitem{}Walker TP, Steigman G, Kang HS, Schramm DM, Olive KA. 1991.
{\it Ap. J.} 376:51--69

\bibitem{}Wang QD, McCray R. 1993.
{\it Ap. J.} 409:L37--40

\bibitem{}Ward MJ, Wilson AS, Penston MV, Elvis M, Maccaccaro T, et al. 1978.
{\it Ap. J.} 223:788--97

\bibitem{}Weiner BJ, Williams TB. 1996.
{\it Astron. J.} 111:1156--63

\bibitem{}Wells A, Abbey AF, Barstow MA, Cole RE, Pye JP, et al. 1990.
{\it Proc. SPIE} 733:519--532

\bibitem{}White DA. 2000.
{\it MNRAS} 312:663--88

\bibitem{}White RA, Bliton M, Bhavsar SP, Bornmann P, Burns JO, et al. 1999.
{\it Astron. J.} 118:2014--37

\bibitem{}White RE. 1991.
{\it Ap. J.} 367:69--77

\bibitem{}White SDM, Navarro JF, Evrard AE, Frenk CS. 1993.
{\it Nature} 366:429--33

\bibitem{}Wu KKS, Fabian AC, Nulsen PEJ. 2000.
{\it MNRAS} in press

\bibitem{}Wu X-P, Xue Y-J, Fang L-Z. 1999.
{\it Ap. J.} 524:22--30

\bibitem{}Xue Y, Wu X-P. 2000.
{\it Ap. J.} 538:65--71

\bibitem{}Zabludoff AI, Mulchaey JS. 1998.
{\it Ap. J.} 496:39--72

\bibitem{}Zamorani G, Mignoli M, Hasinger G, Burg R, Giacconi R, et al. 1999.
{\it Astron. Astrophys.} 346:731--52

\bibitem{}Zwaan MA, Briggs FH. 2000.
{\it Ap. J.} 530:L61--64

\end{thebibliography}
\end{document}